\documentstyle[11pt,aaspp4]{article}
\def\arcs{\ifmmode {^{\scriptscriptstyle\prime\prime}}
          \else $^{\scriptscriptstyle\prime\prime}$\fi}
\def\arcm{\ifmmode {^{\scriptscriptstyle\prime}}
          \else $^{\scriptscriptstyle\prime}$\fi}
\newdimen\sa  \newdimen\sb
\def\parcs{\sa=.07em \sb=.03em
     \ifmmode $\rlap{.}$^{\scriptscriptstyle\prime\kern -\sb\prime}$\kern -\sa$
     \else \rlap{.}$^{\scriptscriptstyle\prime\kern -\sb\prime}$\kern -\sa\fi}
\def\parcm{\sa=.08em \sb=.03em
     \ifmmode $\rlap{.}\kern\sa$^{\scriptscriptstyle\prime}$\kern-\sb$
     \else \rlap{.}\kern\sa$^{\scriptscriptstyle\prime}$\kern-\sb\fi}
\def\pdeg{\ifmmode $\setbox0=\hbox{$^{\circ}$}\rlap{\hskip.11\wd0 .}$^{\circ}
          \else \setbox0=\hbox{$^{\circ}$}\rlap{\hskip.11\wd0 .}$^{\circ}$\fi}
\def\gtorder{\mathrel{\raise.3ex\hbox{$>$}\mkern-14mu
             \lower0.6ex\hbox{$\sim$}}}
\def\ltorder{\mathrel{\raise.3ex\hbox{$<$}\mkern-14mu
             \lower0.6ex\hbox{$\sim$}}}

\newcommand{\kms}{\mbox{ km~s$^{-1}$}}

\def\PsfigVersion{1.10}
\def\setDriver{\DvipsDriver} % \DvipsDriver or \OzTeXDriver
\ifx\undefined\psfig\else\endinput\fi
\let\LaTeXAtSign=\@
\let\@=\relax
\edef\psfigRestoreAt{\catcode`\@=\number\catcode`@\relax}
\catcode`\@=11\relax
\newwrite\@unused
\def\ps@typeout#1{{\let\protect\string\immediate\write\@unused{#1}}}

\def\DvipsDriver{
	\ps@typeout{psfig/tex \PsfigVersion -dvips}
\def\PsfigSpecials{\DvipsSpecials} 	\def\ps@dir{/}
\def\ps@predir{} }
\def\OzTeXDriver{
	\ps@typeout{psfig/tex \PsfigVersion -oztex}
	\def\PsfigSpecials{\OzTeXSpecials}
	\def\ps@dir{:}
	\def\ps@predir{:}
	\catcode`\^^J=5
}

\def\figurepath{./:}

\def\DoPaths#1{\expandafter\EachPath#1\stoplist}
\def\leer{}
\def\EachPath#1:#2\stoplist{% #1 part of the list (delimiter :)
  \ExistsFile{#1}{\SearchedFile}
  \ifx#2\leer
  \else
    \expandafter\EachPath#2\stoplist
  \fi}
\def\ps@dir{/}
\def\ExistsFile#1#2{%
   \openin1=\ps@predir#1\ps@dir#2
   \ifeof1
       \closein1
   \else
       \closein1
        \ifx\ps@founddir\leer
           \edef\ps@founddir{#1}
        \fi
   \fi}
\def\get@dir#1{%
  \def\ps@founddir{}
  \def\SearchedFile{#1}
  \DoPaths\figurepath
}
\def\@nnil{\@nil}
\def\@empty{}
\def\@psdonoop#1\@@#2#3{}
\def\@psdo#1:=#2\do#3{\edef\@psdotmp{#2}\ifx\@psdotmp\@empty \else
    \expandafter\@psdoloop#2,\@nil,\@nil\@@#1{#3}\fi}
\def\@psdoloop#1,#2,#3\@@#4#5{\def#4{#1}\ifx #4\@nnil \else
       #5\def#4{#2}\ifx #4\@nnil \else#5\@ipsdoloop #3\@@#4{#5}\fi\fi}
\def\@ipsdoloop#1,#2\@@#3#4{\def#3{#1}\ifx #3\@nnil 
       \let\@nextwhile=\@psdonoop \else
      #4\relax\let\@nextwhile=\@ipsdoloop\fi\@nextwhile#2\@@#3{#4}}
\def\@tpsdo#1:=#2\do#3{\xdef\@psdotmp{#2}\ifx\@psdotmp\@empty \else
    \@tpsdoloop#2\@nil\@nil\@@#1{#3}\fi}
\def\@tpsdoloop#1#2\@@#3#4{\def#3{#1}\ifx #3\@nnil 
       \let\@nextwhile=\@psdonoop \else
      #4\relax\let\@nextwhile=\@tpsdoloop\fi\@nextwhile#2\@@#3{#4}}
\ifx\undefined\fbox
\newdimen\fboxrule
\newdimen\fboxsep
\newdimen\ps@tempdima
\newbox\ps@tempboxa
\long\def\fbox#1{\leavevmode\setbox\ps@tempboxa\hbox{#1}\ps@tempdima\fboxrule
    \advance\ps@tempdima \fboxsep \advance\ps@tempdima \dp\ps@tempboxa
   \hbox{\lower \ps@tempdima\hbox
  {\vbox{\hrule height \fboxrule
          \hbox{\vrule width \fboxrule \hskip\fboxsep
          \vbox{\vskip\fboxsep \box\ps@tempboxa\vskip\fboxsep}\hskip 
                 \fboxsep\vrule width \fboxrule}
                 \hrule height \fboxrule}}}}
\fi
\newread\ps@stream
\newif\ifnot@eof       % continue looking for the bounding box?
\newif\if@noisy        % report what you're making?
\newif\if@atend        % %%BoundingBox: has (at end) specification
\newif\if@psfile       % does this look like a PostScript file?
{\catcode`\%=12\global\gdef\epsf@start{%!}}
\def\epsf@PS{PS}
\def\epsf@getbb#1{%
\openin\ps@stream=\ps@predir#1
\ifeof\ps@stream\ps@typeout{Error, File #1 not found}\else
   {\not@eoftrue \chardef\other=12
    \def\do##1{\catcode`##1=\other}\dospecials \catcode`\ =10
    \loop
       \if@psfile
	  \read\ps@stream to \epsf@fileline
       \else{
	  \obeyspaces
          \read\ps@stream to \epsf@tmp\global\let\epsf@fileline\epsf@tmp}
       \fi
       \ifeof\ps@stream\not@eoffalse\else
       \if@psfile\else
       \expandafter\epsf@test\epsf@fileline:. \\%
       \fi
          \expandafter\epsf@aux\epsf@fileline:. \\%
       \fi
   \ifnot@eof\repeat
   }\closein\ps@stream\fi}%
\long\def\epsf@test#1#2#3:#4\\{\def\epsf@testit{#1#2}
			\ifx\epsf@testit\epsf@start\else
\ps@typeout{Warning! File does not start with `\epsf@start'.  It may not be a PostScript file.}
			\fi
			\@psfiletrue} % don't test after 1st line
{\catcode`\%=12\global\let\epsf@percent=%\global\def\epsf@bblit{%BoundingBox}}
\long\def\epsf@aux#1#2:#3\\{\ifx#1\epsf@percent
   \def\epsf@testit{#2}\ifx\epsf@testit\epsf@bblit
	\@atendfalse
        \epsf@atend #3 . \\%
	\if@atend	
	   \if@verbose{
		\ps@typeout{psfig: found `(atend)'; continuing search}
	   }\fi
        \else
        \epsf@grab #3 . . . \\%
        \not@eoffalse
        \global\no@bbfalse
        \fi
   \fi\fi}%
\def\epsf@grab #1 #2 #3 #4 #5\\{%
   \global\def\epsf@llx{#1}\ifx\epsf@llx\empty
      \epsf@grab #2 #3 #4 #5 .\\\else
   \global\def\epsf@lly{#2}%
   \global\def\epsf@urx{#3}\global\def\epsf@ury{#4}\fi}%
\def\epsf@atendlit{(atend)} 
\def\epsf@atend #1 #2 #3\\{%
   \def\epsf@tmp{#1}\ifx\epsf@tmp\empty
      \epsf@atend #2 #3 .\\\else
   \ifx\epsf@tmp\epsf@atendlit\@atendtrue\fi\fi}

\chardef\psletter = 11 % won't conflict with \begin{letter} now...
\chardef\other = 12

\newif \ifdebug %%% turn me on to see TeX hard at work ...
\newif\ifc@mpute %%% don't need to compute some values
\c@mputetrue % but assume that we do

\let\then = \relax
\def\r@dian{pt }
\let\r@dians = \r@dian
\let\dimensionless@nit = \r@dian
\let\dimensionless@nits = \dimensionless@nit
\def\internal@nit{sp }
\let\internal@nits = \internal@nit
\newif\ifstillc@nverging
\def \Mess@ge #1{\ifdebug \then \message {#1} \fi}

{ %%% Things that need abnormal catcodes %%%
	\catcode `\@ = \psletter
	\gdef \nodimen {\expandafter \n@dimen \the \dimen}
	\gdef \term #1 #2 #3%
	       {\edef \t@ {\the #1}%%% freeze parameter 1 (count, by value)
		\edef \t@@ {\expandafter \n@dimen \the #2\r@dian}%
		\t@rm {\t@} {\t@@} {#3}%
	       }
	\gdef \t@rm #1 #2 #3%
	       {{%
		\count 0 = 0
		\dimen 0 = 1 \dimensionless@nit
		\dimen 2 = #2\relax
		\Mess@ge {Calculating term #1 of \nodimen 2}%
		\loop
		\ifnum	\count 0 < #1
		\then	\advance \count 0 by 1
			\Mess@ge {Iteration \the \count 0 \space}%
			\Multiply \dimen 0 by {\dimen 2}%
			\Mess@ge {After multiplication, term = \nodimen 0}%
			\Divide \dimen 0 by {\count 0}%
			\Mess@ge {After division, term = \nodimen 0}%
		\repeat
		\Mess@ge {Final value for term #1 of 
				\nodimen 2 \space is \nodimen 0}%
		\xdef \Term {#3 = \nodimen 0 \r@dians}%
		\aftergroup \Term
	       }}
	\catcode `\p = \other
	\catcode `\t = \other
	\gdef \n@dimen #1pt{#1} %%% throw away the ``pt''
}

\def \Divide #1by #2{\divide #1 by #2} %%% just a synonym

\def \Multiply #1by #2%%% allows division of a dimen by a dimen
       {{%%% should really freeze parameter 2 (dimen, passed by value)
	\count 0 = #1\relax
	\count 2 = #2\relax
	\count 4 = 65536
	\Mess@ge {Before scaling, count 0 = \the \count 0 \space and
			count 2 = \the \count 2}%
	\ifnum	\count 0 > 32767 %%% do our best to avoid overflow
	\then	\divide \count 0 by 4
		\divide \count 4 by 4
	\else	\ifnum	\count 0 < -32767
		\then	\divide \count 0 by 4
			\divide \count 4 by 4
		\else
		\fi
	\fi
	\ifnum	\count 2 > 32767 %%% while retaining reasonable accuracy
	\then	\divide \count 2 by 4
		\divide \count 4 by 4
	\else	\ifnum	\count 2 < -32767
		\then	\divide \count 2 by 4
			\divide \count 4 by 4
		\else
		\fi
	\fi
	\multiply \count 0 by \count 2
	\divide \count 0 by \count 4
	\xdef \product {#1 = \the \count 0 \internal@nits}%
	\aftergroup \product
       }}

\def\r@duce{\ifdim\dimen0 > 90\r@dian \then   % sin(x+90) = sin(180-x)
		\multiply\dimen0 by -1
		\advance\dimen0 by 180\r@dian
		\r@duce
	    \else \ifdim\dimen0 < -90\r@dian \then  % sin(-x) = sin(360+x)
		\advance\dimen0 by 360\r@dian
		\r@duce
		\fi
	    \fi}

\def\Sine#1%
       {{%
	\dimen 0 = #1 \r@dian
	\r@duce
	\ifdim\dimen0 = -90\r@dian \then
	   \dimen4 = -1\r@dian
	   \c@mputefalse
	\fi
	\ifdim\dimen0 = 90\r@dian \then
	   \dimen4 = 1\r@dian
	   \c@mputefalse
	\fi
	\ifdim\dimen0 = 0\r@dian \then
	   \dimen4 = 0\r@dian
	   \c@mputefalse
	\fi
	\ifc@mpute \then
		\divide\dimen0 by 180
		\dimen0=3.141592654\dimen0
		\dimen 2 = 3.1415926535897963\r@dian %%% a well-known constant
		\divide\dimen 2 by 2 %%% we only deal with -pi/2 : pi/2
		\Mess@ge {Sin: calculating Sin of \nodimen 0}%
		\count 0 = 1 %%% see power-series expansion for sine
		\dimen 2 = 1 \r@dian %%% ditto
		\dimen 4 = 0 \r@dian %%% ditto
		\loop
			\ifnum	\dimen 2 = 0 %%% then we've done
			\then	\stillc@nvergingfalse 
			\else	\stillc@nvergingtrue
			\fi
			\ifstillc@nverging %%% then calculate next term
			\then	\term {\count 0} {\dimen 0} {\dimen 2}%
				\advance \count 0 by 2
				\count 2 = \count 0
				\divide \count 2 by 2
				\ifodd	\count 2 %%% signs alternate
				\then	\advance \dimen 4 by \dimen 2
				\else	\advance \dimen 4 by -\dimen 2
				\fi
		\repeat
	\fi		
			\xdef \sine {\nodimen 4}%
       }}

\def\Cosine#1{\ifx\sine\UnDefined\edef\Savesine{\relax}\else
		             \edef\Savesine{\sine}\fi
	{\dimen0=#1\r@dian\advance\dimen0 by 90\r@dian
	 \Sine{\nodimen 0}
	 \xdef\cosine{\sine}
	 \xdef\sine{\Savesine}}}	      
\def\psdraft{
	\def\@psdraft{0}
}
\def\psfull{
	\def\@psdraft{100}
}

\psfull

\newif\if@scalefirst
\def\psscalefirst{\@scalefirsttrue}
\def\psrotatefirst{\@scalefirstfalse}
\psrotatefirst

\newif\if@draftbox
\def\psnodraftbox{
	\@draftboxfalse
}
\def\psdraftbox{
	\@draftboxtrue
}
\@draftboxtrue

\newif\if@prologfile
\newif\if@postlogfile
\def\pssilent{
	\@noisyfalse
}
\def\psnoisy{
	\@noisytrue
}
\psnoisy
\newif\if@bbllx
\newif\if@bblly
\newif\if@bburx
\newif\if@bbury
\newif\if@height
\newif\if@width
\newif\if@rheight
\newif\if@rwidth
\newif\if@angle
\newif\if@clip
\newif\if@verbose
\def\@p@@sclip#1{\@cliptrue}
\newif\if@decmpr
\def\@p@@sfigure#1{\def\@p@sfile{null}\def\@p@sbbfile{null}\@decmprfalse
   \openin1=\ps@predir#1
   \ifeof1
	\closein1
	\get@dir{#1}
	\ifx\ps@founddir\leer
		\openin1=\ps@predir#1.bb
		\ifeof1
			\closein1
			\get@dir{#1.bb}
			\ifx\ps@founddir\leer
				\ps@typeout{Can't find #1 in \figurepath}
			\else
				\@decmprtrue
				\def\@p@sfile{\ps@founddir\ps@dir#1}
				\def\@p@sbbfile{\ps@founddir\ps@dir#1.bb}
			\fi
		\else
			\closein1
			\@decmprtrue
			\def\@p@sfile{#1}
			\def\@p@sbbfile{#1.bb}
		\fi
	\else
		\def\@p@sfile{\ps@founddir\ps@dir#1}
		\def\@p@sbbfile{\ps@founddir\ps@dir#1}
	\fi
   \else
	\closein1
	\def\@p@sfile{#1}
	\def\@p@sbbfile{#1}
   \fi
}
\def\@p@@sfile#1{\@p@@sfigure{#1}}
\def\@p@@sbbllx#1{
		\@bbllxtrue
		\dimen100=#1
		\edef\@p@sbbllx{\number\dimen100}
}
\def\@p@@sbblly#1{
		\@bbllytrue
		\dimen100=#1
		\edef\@p@sbblly{\number\dimen100}
}
\def\@p@@sbburx#1{
		\@bburxtrue
		\dimen100=#1
		\edef\@p@sbburx{\number\dimen100}
}
\def\@p@@sbbury#1{
		\@bburytrue
		\dimen100=#1
		\edef\@p@sbbury{\number\dimen100}
}
\def\@p@@sheight#1{
		\@heighttrue
		\dimen100=#1
   		\edef\@p@sheight{\number\dimen100}
}
\def\@p@@swidth#1{
		\@widthtrue
		\dimen100=#1
		\edef\@p@swidth{\number\dimen100}
}
\def\@p@@srheight#1{
		\@rheighttrue
		\dimen100=#1
		\edef\@p@srheight{\number\dimen100}
}
\def\@p@@srwidth#1{
		\@rwidthtrue
		\dimen100=#1
		\edef\@p@srwidth{\number\dimen100}
}
\def\@p@@sangle#1{
		\@angletrue
		\edef\@p@sangle{#1} %\number\dimen100}
}
\def\@p@@ssilent#1{ 
		\@verbosefalse
}
\def\@p@@sprolog#1{\@prologfiletrue\def\@prologfileval{#1}}
\def\@p@@spostlog#1{\@postlogfiletrue\def\@postlogfileval{#1}}
\def\@cs@name#1{\csname #1\endcsname}
\def\@setparms#1=#2,{\@cs@name{@p@@s#1}{#2}}
\def\ps@init@parms{
		\@bbllxfalse \@bbllyfalse
		\@bburxfalse \@bburyfalse
		\@heightfalse \@widthfalse
		\@rheightfalse \@rwidthfalse
		\def\@p@sbbllx{}\def\@p@sbblly{}
		\def\@p@sbburx{}\def\@p@sbbury{}
		\def\@p@sheight{}\def\@p@swidth{}
		\def\@p@srheight{}\def\@p@srwidth{}
		\def\@p@sangle{0}
		\def\@p@sfile{} \def\@p@sbbfile{}
		\def\@p@scost{10}
		\def\@sc{}
		\@prologfilefalse
		\@postlogfilefalse
		\@clipfalse
		\if@noisy
			\@verbosetrue
		\else
			\@verbosefalse
		\fi
}
\def\parse@ps@parms#1{
	 	\@psdo\@psfiga:=#1\do
		   {\expandafter\@setparms\@psfiga,}}
\newif\ifno@bb
\def\bb@missing{
	\if@verbose{
		\ps@typeout{psfig: searching \@p@sbbfile \space  for bounding box}
	}\fi
	\no@bbtrue
	\epsf@getbb{\@p@sbbfile}
        \ifno@bb \else \bb@cull\epsf@llx\epsf@lly\epsf@urx\epsf@ury\fi
}	
\def\bb@cull#1#2#3#4{
	\dimen100=#1 bp\edef\@p@sbbllx{\number\dimen100}
	\dimen100=#2 bp\edef\@p@sbblly{\number\dimen100}
	\dimen100=#3 bp\edef\@p@sbburx{\number\dimen100}
	\dimen100=#4 bp\edef\@p@sbbury{\number\dimen100}
	\no@bbfalse
}
\newdimen\p@intvaluex
\newdimen\p@intvaluey
\def\rotate@#1#2{{\dimen0=#1 sp\dimen1=#2 sp
		  \global\p@intvaluex=\cosine\dimen0
		  \dimen3=\sine\dimen1
		  \global\advance\p@intvaluex by -\dimen3
		  \global\p@intvaluey=\sine\dimen0
		  \dimen3=\cosine\dimen1
		  \global\advance\p@intvaluey by \dimen3
		  }}
\def\compute@bb{
		\no@bbfalse
		\if@bbllx \else \no@bbtrue \fi
		\if@bblly \else \no@bbtrue \fi
		\if@bburx \else \no@bbtrue \fi
		\if@bbury \else \no@bbtrue \fi
		\ifno@bb \bb@missing \fi
		\ifno@bb \ps@typeout{FATAL ERROR: no bb supplied or found}
			\no-bb-error
		\fi
		\count203=\@p@sbburx
		\count204=\@p@sbbury
		\advance\count203 by -\@p@sbbllx
		\advance\count204 by -\@p@sbblly
		\edef\ps@bbw{\number\count203}
		\edef\ps@bbh{\number\count204}
		\if@angle 
			\Sine{\@p@sangle}\Cosine{\@p@sangle}
	        	{\dimen100=\maxdimen\xdef\r@p@sbbllx{\number\dimen100}
					    \xdef\r@p@sbblly{\number\dimen100}
			                    \xdef\r@p@sbburx{-\number\dimen100}
					    \xdef\r@p@sbbury{-\number\dimen100}}
                        \def\minmaxtest{
			   \ifnum\number\p@intvaluex<\r@p@sbbllx
			      \xdef\r@p@sbbllx{\number\p@intvaluex}\fi
			   \ifnum\number\p@intvaluex>\r@p@sbburx
			      \xdef\r@p@sbburx{\number\p@intvaluex}\fi
			   \ifnum\number\p@intvaluey<\r@p@sbblly
			      \xdef\r@p@sbblly{\number\p@intvaluey}\fi
			   \ifnum\number\p@intvaluey>\r@p@sbbury
			      \xdef\r@p@sbbury{\number\p@intvaluey}\fi
			   }
			\rotate@{\@p@sbbllx}{\@p@sbblly}
			\minmaxtest
			\rotate@{\@p@sbbllx}{\@p@sbbury}
			\minmaxtest
			\rotate@{\@p@sbburx}{\@p@sbblly}
			\minmaxtest
			\rotate@{\@p@sbburx}{\@p@sbbury}
			\minmaxtest
			\edef\@p@sbbllx{\r@p@sbbllx}\edef\@p@sbblly{\r@p@sbblly}
			\edef\@p@sbburx{\r@p@sbburx}\edef\@p@sbbury{\r@p@sbbury}
		\fi
		\count203=\@p@sbburx
		\count204=\@p@sbbury
		\advance\count203 by -\@p@sbbllx
		\advance\count204 by -\@p@sbblly
		\edef\@bbw{\number\count203}
		\edef\@bbh{\number\count204}
}
\def\in@hundreds#1#2#3{\count240=#2 \count241=#3
		     \count100=\count240	% 100 is first digit #2/#3
		     \divide\count100 by \count241
		     \count101=\count100
		     \multiply\count101 by \count241
		     \advance\count240 by -\count101
		     \multiply\count240 by 10
		     \count101=\count240	%101 is second digit of #2/#3
		     \divide\count101 by \count241
		     \count102=\count101
		     \multiply\count102 by \count241
		     \advance\count240 by -\count102
		     \multiply\count240 by 10
		     \count102=\count240	% 102 is the third digit
		     \divide\count102 by \count241
		     \count200=#1\count205=0
		     \count201=\count200
			\multiply\count201 by \count100
		 	\advance\count205 by \count201
		     \count201=\count200
			\divide\count201 by 10
			\multiply\count201 by \count101
			\advance\count205 by \count201
		     \count201=\count200
			\divide\count201 by 100
			\multiply\count201 by \count102
			\advance\count205 by \count201
		     \edef\@result{\number\count205}
}
\def\compute@wfromh{
		\in@hundreds{\@p@sheight}{\@bbw}{\@bbh}
		\edef\@p@swidth{\@result}
}
\def\compute@hfromw{
	        \in@hundreds{\@p@swidth}{\@bbh}{\@bbw}
		\edef\@p@sheight{\@result}
}
\def\compute@handw{
		\if@height 
			\if@width
			\else
				\compute@wfromh
			\fi
		\else 
			\if@width
				\compute@hfromw
			\else
				\edef\@p@sheight{\@bbh}
				\edef\@p@swidth{\@bbw}
			\fi
		\fi
}
\def\compute@resv{
		\if@rheight \else \edef\@p@srheight{\@p@sheight} \fi
		\if@rwidth \else \edef\@p@srwidth{\@p@swidth} \fi
}
\def\compute@sizes{
	\compute@bb
	\if@scalefirst\if@angle
	\if@width
	   \in@hundreds{\@p@swidth}{\@bbw}{\ps@bbw}
	   \edef\@p@swidth{\@result}
	\fi
	\if@height
	   \in@hundreds{\@p@sheight}{\@bbh}{\ps@bbh}
	   \edef\@p@sheight{\@result}
	\fi
	\fi\fi
	\compute@handw
	\compute@resv}
\def\OzTeXSpecials{
	\special{empty.ps /@isp {true} def}
	\special{empty.ps \@p@swidth \space \@p@sheight \space
			\@p@sbbllx \space \@p@sbblly \space
			\@p@sbburx \space \@p@sbbury \space
			startTexFig \space }
	\if@clip{
		\if@verbose{
			\ps@typeout{(clip)}
		}\fi
		\special{empty.ps doclip \space }
	}\fi
	\if@angle{
		\if@verbose{
			\ps@typeout{(rotate)}
		}\fi
		\special {empty.ps \@p@sangle \space rotate \space} 
	}\fi
	\if@prologfile
	    \special{\@prologfileval \space } \fi
	\if@decmpr{
		\if@verbose{
			\ps@typeout{psfig: Compression not available
			in OzTeX version \space }
		}\fi
	}\else{
		\if@verbose{
			\ps@typeout{psfig: including \@p@sfile \space }
		}\fi
		\special{epsf=\ps@predir\@p@sfile \space }
	}\fi
	\if@postlogfile
	    \special{\@postlogfileval \space } \fi
	\special{empty.ps /@isp {false} def}
}
\def\DvipsSpecials{
	\special{ps::[begin] 	\@p@swidth \space \@p@sheight \space
			\@p@sbbllx \space \@p@sbblly \space
			\@p@sbburx \space \@p@sbbury \space
			startTexFig \space }
	\if@clip{
		\if@verbose{
			\ps@typeout{(clip)}
		}\fi
		\special{ps:: doclip \space }
	}\fi
	\if@angle
		\if@verbose{
			\ps@typeout{(clip)}
		}\fi
		\special {ps:: \@p@sangle \space rotate \space} 
	\fi
	\if@prologfile
	    \special{ps: plotfile \@prologfileval \space } \fi
	\if@decmpr{
		\if@verbose{
			\ps@typeout{psfig: including \@p@sfile.Z \space }
		}\fi
		\special{ps: plotfile "`zcat \@p@sfile.Z" \space }
	}\else{
		\if@verbose{
			\ps@typeout{psfig: including \@p@sfile \space }
		}\fi
		\special{ps: plotfile \@p@sfile \space }
	}\fi
	\if@postlogfile
	    \special{ps: plotfile \@postlogfileval \space } \fi
	\special{ps::[end] endTexFig \space }
}
\def\psfig#1{\vbox {
	\ps@init@parms
	\parse@ps@parms{#1}
	\compute@sizes
	\ifnum\@p@scost<\@psdraft{
		\PsfigSpecials 
		\vbox to \@p@srheight sp{
			\hbox to \@p@srwidth sp{
				\hss
			}
		\vss
		}
	}\else{
		\if@draftbox{		
			\hbox{\fbox{\vbox to \@p@srheight sp{
			\vss
			\hbox to \@p@srwidth sp{ \hss 
			 \hss }
			\vss
			}}}
		}\else{
			\vbox to \@p@srheight sp{
			\vss
			\hbox to \@p@srwidth sp{\hss}
			\vss
			}
		}\fi

	}\fi
}}
\psfigRestoreAt
\setDriver
\let\@=\LaTeXAtSign

\lefthead{Kochanek et al.}
\righthead{}

\def\ntot{$41$}
\def\nuse{$30$}
\def\nclst{$50$}
\def\ndrop{$11$}
\def\nsrcz{$24$}
\def\nlenz{$20$}
\def\nosrcz{$ 6$}
\def\nolenz{$10$}
\def\clopt{$-0.03 \pm  0.11$}
\def\lnopt{$-0.05 \pm  0.22$}
\def\lnir{$-0.02 \pm  0.25$}
\def\lnzwsrc{-0.05 \pm  0.09}
\def\lnzall{-0.04 \pm  0.09}
\def\clzall{-0.03 \pm  0.06}

\begin{document}

\title{The Fundamental Plane of Gravitational Lens Galaxies \\ and \\
  The Evolution of Early-Type Galaxies in Low Density Environments \footnote{Based on 
Observations made with the NASA/ESA Hubble Space
Telescope, obtained at the Space Telescope Science Institute, which is
operated by AURA, Inc., under NASA contract NAS 5-26555. }} 

\vskip 2truecm

\author{C.S. Kochanek$^{(a)}$,} 
\author{E.E. Falco$^{(a)}$, C.D. Impey$^{(b)}$, J. Leh\'ar$^{(a)}$, 
B.A. McLeod$^{(a)}$}

\author{H.-W. Rix$^{(c)}$, C.R. Keeton$^{(b)}$, J.A. Mu\~noz$^{(a)}$ and C.Y. Peng$^{(b)}$}

\affil{$^{(a)}$ Harvard-Smithsonian Center for Astrophysics, 60 Garden
	St., Cambridge, MA 02138}
\affil{email: ckochanek, efalco, jlehar, bmcleod, jmunoz@cfa.harvard.edu}
\affil{$^{(b)}$Steward Observatory, University of Arizona, Tucson, AZ 85721}
\affil{email: impey,  ckeeton, cyp@as.arizona.edu}
\affil{$^{(c)}$Max-Planck-Institut fuer Astronomie, Koenigsstuhl 17, D-69117 Heidelberg, Germany}
\affil{email: rix@mpia-hd.mpg.de}

\begin{abstract}
Most gravitational lenses are early-type galaxies in relatively low density 
environments -- a ``field'' rather than a ``cluster'' population.   Their
average properties are the {\it mass-averaged} properties of all early-type 
galaxies.
We show that field early-type galaxies with $0 < z < 1$, as represented by the 
lens galaxies, lie on the same fundamental plane as those in rich clusters
at similar redshifts.  We then use the fundamental plane to measure the 
combined evolutionary and K-corrections for early-type galaxies in the V, I 
and H bands.  Only for passively evolving stellar populations formed at 
$z_f \gtorder 2$ ($H_0=65$~km~s$^{-1}$~Mpc$^{-1}$, $\Omega_0=0.3$, $\lambda_0=0.7$) 
can the lens galaxies be matched to the local fundamental plane. 
  The high formation epoch and the lack of significant 
differences between the field and cluster populations contradict many
current models of the formation history of early-type galaxies.  Lens
galaxy colors and the fundamental plane provide good photometric redshift
estimates with an empirical accuracy of
$\langle z_{FP} - z_l \rangle = \lnzall$ for the \nlenz$\,$ lenses with 
known redshifts.  A mass model dominated by dark matter is more consistent
with the data than either an isotropic or radially anisotropic constant
M/L mass model, and a radially anisotropic model is better
than an isotropic model. 
\end{abstract}


\section{Introduction}

The formation and evolution of galaxies is a central problem of modern astronomy.  
In particular, observations show that most early-type galaxies in rich clusters formed 
their stars at an early epoch ($z_f\sim 2$--$3$) and have only evolved passively 
during the following 10~Gyr (e.g. Bower, Lucey \& Ellis 1992).  The early-type galaxies 
in clusters have extraordinarily uniform colors both internally and among clusters, 
as well as very tight correlations between color and velocity dispersion that are 
difficult to reconcile with a wide range of ages for their stellar populations.
The uniformity of the colors persists to $z\sim 1$ (Ellis et al. 1997, Stanford, 
Eisenhardt \& Dickinson 1998, Pahre 1999), 
although there is evidence that the S0 galaxies are evolving 
faster than the ellipticals (van Dokkum et al. 1998b).  The fundamental plane
or FP (Djorgovski \& Davis 1987, Dressler et al. 1987), a tight correlation
between effective radius, surface brightness and velocity dispersion for early-type
galaxies, provides a powerful tool for probing their evolution.  Local measurements 
of the FP at a range of wavelengths (e.g. Jorgensen, Franx \& Kjaergaard 1995ab, 1996, 
Pahre, de Carvalho \& Djorgovski 1998a) can be combined with measurements of
the FP in rich clusters at intermediate redshifts (van Dokkum \& Franx 1996,
Kelson et al. 1997, 2000, van Dokkum et al. 1998a, Pahre, Djorgovski \& de Carvalho 1999ab, 
Jorgensen et al.  1999) to directly measure the evolution of the mass-to-light ratio 
of early-type galaxies with cosmic epoch.  This evolution is consistent with an early 
formation epoch for the stellar populations, and the FP results probably rule out,
at least for cluster galaxies,  the broad range of formation epochs inferred from modeling 
the line strengths of local early-type galaxies (e.g.  Trager 1997, Jorgensen 1999, 
Terlevich et al. 1999, see Pahre, de Carvalho \& Djorgovski 1998b).  
 
Far less is known about the homogeneity and evolution of early-type galaxies in
less dense environments than the cores of rich clusters, even though these 
galaxies represent the vast majority of the early-type galaxies. Semi-analytic models 
of galaxy formation, particularly 
those of Kauffmann (1996) and Kauffmann \& Charlot (1998), predict that field early-type 
galaxies have very late forming stellar populations ($z_f < 1$), while cluster 
early-types have significantly older populations.  In their models, however, the 
halos they identify as early-type galaxies appear to have old stellar populations 
at all epochs because the models also predict a rapidly evolving number density of 
early-type galaxies.  The preponderance of the observational evidence suggests, however, 
that there is little evolution in the number of massive early-type galaxies to $z\simeq1$ (Lilly et al.
1995, Schade et al. 1999), although contrary views exist (Kauffmann, Charlot \& White 1996).    
If the number density evolves little, then the early-type galaxies near $z=1$
must be the precursors of those at $z=0$.  Studies of local early-type galaxies
find some evidence that field early-type galaxies have younger stellar populations (e.g. de Carvalho 
\& Djorgovski 1992, Guzm\'an \& Lacey 1993, Forbes, Ponman \& Brown 1998, James \& 
Mobasher 1999), but with similar difficulties untangling age from metallicity as 
are found in the cluster samples.  As with the cluster galaxies, the best way to
separate age from metallicity is to look at earlier epochs.  Schade et al. (1996, 1999) 
have used the correlation between effective radius and luminosity to show that the
luminosity evolution of field and cluster early-type galaxies is similar at 
$z\sim 0.5$. Treu et al. (1999) constructed the fundamental plane of six field 
early-types near $z=0.3$ and found that it was consistent with that found
by van Dokkum \& Franx (1996) and Kelson et al. (1997) for the cluster sample.
The absence of extremely red galaxies in deep surveys sets a weak upper bound on the
star formation epoch of $z_f\lesssim 5$ (e.g. Zepf 1997).  It is a weak upper
bound because a very small amount of late-time star formation will make a galaxy
bluer than the extreme colors ($V-K>7$ mag) used to obtain the limit.

The population of gravitational lens galaxies is dominated by massive early-type galaxies
(Keeton, Kochanek \& Falco 1998), as expected from theoretical predictions (e.g. Fukugita \& 
Turner 1991, Maoz \& Rix 1993, Kochanek 1993, 1996).  The lens galaxies are selected based on their
mass rather than on any property related to star formation,\footnote{The optically selected
lenses are biased against star forming lens galaxies to the extent that any associated 
dust obscures background sources (see Falco et al. 1999).  Because almost all optically-selected 
lenses are very bright quasars, the lower mass-to-light ratios of star forming
galaxies do not produce a bias (see Kochanek 1996).  Radio selected lenses are immune to both effects.} 
leading to a sample dominated by $L_*$ early-type galaxies.  Indeed, since the lensing cross
section is closely related to the mass of the lens, the average properties of the lens galaxies
at any redshift are nearly identical to the {\it mass-weighted} average properties of all galaxies
at that redshift.  The lens galaxies are also a ``fair'' sample of the environmental distribution 
of early-type galaxies.  The exception is that galaxy-dominated lenses will not be found in the 
cores of rich clusters where the cluster potential dominates any lensing effects and produces
phenomenon such as giant arcs rather than the lenses we discuss here.  Thus, we
will refer to the lenses as a sample of ``field'' early-type galaxies or early-type galaxies
in ``low-density'' environments, as very few lie in group or cluster potentials with velocity 
dispersions larger than the $400\kms$ break used by Kauffmann \& Charlot (1998) to divide early-type 
galaxies into field and cluster samples. 
Keeton et al. (1998) estimated the evolution of the mass-to-light ratio
of the lens galaxies with redshift, and found rates strikingly similar to those
measured for the rich cluster samples.

The CASTLES (CfA-Arizona Space Telescope Lens Survey) survey is obtaining 
V, I and H photometry of the $60$ known lens systems.  Since the geometry
of a gravitational lens provides an accurate measurement of the lens galaxy's 
mass, the lens galaxies are one of the largest samples of galaxies with accurately
measured masses at intermediate redshifts, and they are by far the largest such 
sample outside the very special environments represented by the cores of rich clusters.  
In \S2 we describe our analysis methods and the
data. In \S3 we compare the colors of the lens galaxies to those of the rich
cluster galaxies used in the FP studies at intermediate redshift.  In \S4
we show that the lens galaxies lie on the same FP as the cluster galaxies
at comparable redshifts.  In \S5 we explore photometric redshift estimates for
the lens galaxies, and in \S6 we explore the dark matter problem in early-type
galaxies.  In \S7 we use the FP to measure the evolution of early-type 
galaxies in the V, I and H bands as a function of redshift.  Finally, in \S8
we summarize our results.

\section{Methods and Models}

In this section we detail our local comparison sample (\S2.1) and our method for
using the FP to study the evolution of individual galaxies (\S2.2).  Next we 
discuss how we estimate stellar velocity dispersions from the lensed image separations 
in several dynamical models (\S2.3) and the available sample of gravitational
lens galaxies (\S2.4).  Finally, we discuss the interpretation of the results
using either spectrophotometric models (\S2.5) or comparisons to early-type galaxies
in clusters at comparable redshifts (\S2.6).  

\subsection{The Local FP}

Unlike the studies of the FP in rich clusters, we must work with individual galaxies
spread over a wide range of redshifts.  Thus our analysis methods must be designed to
interpret the data on individual galaxies rather than ensembles of galaxies at a 
common redshift. We use the early-type galaxies in nine local clusters studied
by Jorgensen \& Franx (1994) and Jorgensen, Franx \& Kjaergaard (1995ab, 1996,
collectively referred to as JFK hereafter) as a local comparison sample.
We arbitrarily renormalized the JFK FP to the closest standard HST
filter, F606W, from Gunn r assuming a constant color of $F606W-r=-0.09$ mag
for early-type galaxies (Fukugita, Shimasaku \& Ichikawa 1995).  We also converted
from the angular effective radius of the galaxies at the redshift of Coma 
($cz\equiv7200\kms$) to a physical effective radius for 
$H_0=50 h_{50}$ \kms~Mpc$^{-1}$.  We then redetermined the zero points of the 
FP relations holding the slopes fixed to the JFK values.  For example, the
physical effective radius $r_e^{FP}$ predicted by the fundamental plane 
from the central velocity dispersion and the surface brightness of the galaxy 
is
\begin{equation}
   \log \left({ r_e^{FP} \over h_{50}^{-1}\hbox{kpc} } \right) =  
  1.24\log \left( { \sigma_c \over \kms}\right) + 0.33 \left( { \mu_e(0)\over \hbox{mag/arcsec}^2 } \right) - 8.66,
\end{equation}
with a dispersion of $0.07$ dex (90\% of the galaxies lie within $\pm0.12$ dex).  
The velocity dispersion $\sigma_c$ is 
measured in a fixed physical aperture defined by an angular aperture 3\farcs4 
in diameter at Coma (Jorgensen et al. 1995b), the intermediate axis effective radius is $r_e$, and 
the mean surface brightness ($\mu_e = m + 5 \log r_e + 2.5\log 2\pi$) is 
corrected to zero redshift (Jorgensen \& Franx 1994). 
When analyzing individual galaxies it is useful
to view the FP as a means of predicting the surface brightness the galaxy 
will have at zero redshift.  If we again hold the slopes 
fixed, we find that the FP relation for the zero-redshift surface brightness is
\begin{equation}
   \left( { \mu_{F606W}^{FP}(0) \over \hbox{mag/arcsec}^2 } \right) =  
   -3.76\log \left({ \sigma_c \over \kms }\right) + 
   3.03 \log \left({ r_e \over h_{50}^{-1} \hbox{kpc}} \right) + 26.25 
\end{equation}
with a scatter of $0.23$ mag/arcsec$^{-2}$ in the local JFK sample (90\% of the galaxies
lie within $\pm0.37$ mag/arcsec$^{-2}$).  The covariances of $\log r_e$ and $\mu_e$ make
the variable combination $\log r_e - 0.33\mu_e$ appearing in the expressions for the FP 
insensitive to measurement errors, with uncertainties 5--10 times smaller than
those in either $\log r_e$ or $\mu_e$ alone (see Jorgensen, Franx \& Kjaergaard 1993).

\subsection{Using the FP to Measure Galaxy Evolution}

We will use the FP to study the surface brightness evolution of early-type 
galaxies. We observe galaxies through a series of filters $j$ from which we can
compute the surface brightness $\mu_j(z)$.  The surface brightness evolves as
\begin{equation}
   \mu_j(z) = \mu_j(0) + 10\log (1+z) + e_j(z) + k_j(z),
\end{equation}
where the first redshift term is the $(1+z)^4$ cosmological dimming, the second
is the evolution correction for filter $j$ in the galaxy rest frame, and the 
third term is the K-correction from the rest frame to the observed frame.  
We estimate $\mu_j(0)$ using the FP (eqn. 2) and the
angular diameter distance $D_A(z)$ to the galaxy.
The difference between the observed surface brightness and the estimated 
zero-redshift surface brightness, $\mu^{FP}_j(0)$, is a direct measurement of the    
evolution and K-correction terms for band $j$, 
\begin{equation} 
   e_j(z)+k_j(z) = \mu_j(z) - \left(\mu^{FP}_j(0)+ 10\log(1+z) \right).
\end{equation}
We will not interpolate between filters to eliminate the K-correction 
(as is done for most of the cluster studies) because our filter coverage
is incomplete. A spectrophotometric model is needed to interpret
the results in either case, so the only advantage to measuring $e_j(z)$ or
the mass-to-light ratio at a fixed rest wavelength is pedagogic. 
The values of $e_j$ and $k_j$ depend on the cosmological model only through the 
angular diameter distance used to estimate the physical effective radius (and the velocity
dispersion for the gravitational lenses).  We will not, at present, include a 
model for the modest changes in the slope of the FP with wavelength because there 
is still considerable uncertainty in its measurement (see Pahre, de Carvalho \& Djorgovski 1998ab).
 
\subsection{Lensed Image Separations and Stellar Velocity Dispersions}

For almost all lenses we need to estimate the central velocity dispersion
$\sigma_c$ from the geometry of the lensed images rather than using 
direct measurements.  The image separation $\Delta\theta$
accurately determines the mass of the lens on that scale, which we
must convert into an estimate of $\sigma_c$.\footnote{
The image separation $\Delta\theta$ is defined to be twice the critical
radius of the best fitting singular isothermal sphere (SIS) in an external
tidal shear model for the image positions using a simple source plane
$\chi^2$. This definition is closely related to the mass enclosed by
the images and roughly equal to the average distance of the lensed
images from the lens galaxy.  Using the maximum image separation leads
to larger residuals in the FP. } The best explored model
is the dark matter or singular isothermal sphere (SIS) mass model.
In the SIS model the image separation, $\Delta\theta=8\pi(\sigma_D/c)^2 D_{LS}/D_{OS}$
where $D_{LS}$ and $D_{OS}$ are the angular diameter distances from the lens to the source 
and from the observer to the source respectively, is set by the velocity dispersion
of the dark matter $\sigma_D$ rather than the central velocity dispersion of
the stars $\sigma_c$.  Dynamical models of nearby early-type galaxies in SIS halos 
by Kochanek (1994) demonstrated that the two have nearly identical values,  with
$\langle \sigma_c -\sigma_D \rangle = (8\pm 26) $ km s$^{-1}$.  Models for the
observed distribution of image separations (Kochanek 1996, Falco et al. 1998)
yield $\sigma_{c*} \simeq \sigma_{D*} \simeq (225\pm25)\kms$ for the velocity 
dispersion of an $L_*$ galaxy.  Thus, we estimate the stellar velocity dispersion by 
\begin{equation}
    \sigma_c = { 225 \over f } 
\left( { \Delta \theta \over  2\farcs91 } { D_{OS} \over D_{LS}} \right)^{1/2}
  \hbox{km s}^{-1}.
\end{equation}
which is normalized for an $L_*$ galaxy.  We include a factor $f = \sigma_{D}/\sigma_{c} \simeq 1.0 \pm 0.1$
for the uncertainties in the dynamical normalization.\footnote{For the historically minded, the factor
$f$ is identical to the correction factor originally introduced by Turner, Ostriker
\& Gott (1984), who advocated a value of $f=(3/2)^{1/2}=1.22$ based on simple models of luminous
galaxies in dark matter halos.  A normalization this high is inconsistent with both the dynamics
of nearby galaxies and the observed distribution of image separations. } 
Measurement errors in $\Delta\theta$ are well under 5\% even in the worst cases, and can be
ignored compared to the systematic errors in converting $\Delta\theta$ into
a velocity dispersion.  Note that a 14\% change in $f$ is needed
before the shift in the FP matches its local thickness.   It is 
important to remember that the dark matter model, despite its simplicity, is not
used merely for analytic convenience.  It is the mass model that is most 
consistent with the lensing data (e.g. Kochanek 1995), local stellar dynamics of ellipticals 
(e.g. Rix et al. 1997) and the X-ray halos of ellipticals (e.g. Fabbiano 1989).

We also experimented with constant $M/L$ dynamical models even though they are less
physically plausible.  We assumed the 
mass and luminosity distribution could be approximated by a Hernquist (1990) 
model, and computed constant isotropy parameter $\beta$ (Binney \& Mamon (1982),
$\beta=0$ is isotropic, $\beta=1$ is purely radial) stellar dynamical models normalized to have 
the observed mass inside the Einstein ring of the lens.  For Hernquist radius
$a$ and total mass $M_T$, the central velocity dispersion averaged over 
aperture $R_v$ can be written as $\sigma_c^2 = (GM_T/a) f_1(R_v/a)$ where 
$f_1(R_v/a)$ is a dimensionless function of the dynamical model.  The lensing 
mass of $M_L = \pi \Delta \theta^2 \Sigma_{crit}/4$ is independent of the
dynamical model and the mass distribution and it is related to the total mass 
by $M_L = M_T f_0 (\Delta\theta/2a)$ where $f_0(\Delta\theta/2a)$ is the 
fraction of the total mass inside projected radius $\Delta\theta/2$.  Thus, the 
central velocity dispersion is determined by
\begin{equation}
       \sigma_c^2 = 4 \pi \left( { c \over 7200} \right)^2 
     \left( { \Delta \theta \over 1\farcs0 } \right)^2 \left( { 1\farcs0 \over a } \right) { D_{OS} \over D_{LS} }
                       { f_1 (R_v/a) \over f_0 (\Delta\theta/2a) }.
\end{equation}
The expression is independent of $H_0$ because the distance dependent terms 
appear only in ratios.  We adopt the standard dynamical aperture of
$2R_v=3\farcs4$ at the Coma cluster from JFK.  We set the Hernquist radius 
to $a= 0.551 r_e$ (Hernquist 1990) and we will consider both isotropic
($\beta=0$) and fairly radial ($\beta=0.5$) dynamical models with solutions
determined using the Jeans equations.  The isotropy
assumptions are restrictive, and more sophisticated models allow considerable
freedom in the central velocity dispersion (see Romanowsky \& Kochanek 1999). 

\subsection{HST Photometry of Lens Galaxies}

We considered a sample of \nuse$\,$ lenses observed with either WFPC2 or
NICMOS in the V=F555W, F606W, R=F675W, F702W, F791W, I=F814W, J=F110W, H=F160W and
K=F205W filters.  A summary of the observations is presented in Table 1.  
Of these \nuse$\,$ systems, \nolenz$\,$ are missing spectroscopic
lens redshifts and \nosrcz$\,$ are missing spectroscopic source redshifts.
We regard the V, I and H filters as our standards and we
will frequently use spectrophotometric models to convert measurements in the
other filters to the standard filters.  We assumed Vega-normalized
zero-points (1~count/s
in an infinite aperture) of 21.88, 21.80 and 22.47 mag for the F205W, F160W, 
and F110W NIC2 filters, 21.51 for the F160W NIC1 filter (see Leh\'ar et al. 2000), 
and 21.69, 21.57, 22.47, 22.08, 22.93
and 22.57 mag for the F814W, F791W, F702W, F675W, F606W and F555W filters for a 
gain of 7 (Holtzman et al. 1995, corrected to infinite aperture).
Our full data reduction and analysis procedures are detailed
in Leh\'ar et al. (2000).  For each lens galaxy we determined the 
intermediate axis effective radius on the image having the best
signal-to-noise for the galaxy.  With the scale and shape of the lens galaxy
fixed, we determined the surface brightnesses for the remaining filters.
The uncertainties were estimated as the quadrature sum of a statistical term 
from separate fits to the individual subexposures, a PSF uncertainty term 
from fits with multiple PSFs, and a modeling uncertainty term from the 
difference between two independent modeling codes.  The uncertainties were
separately recorded for colors, magnitudes, and the variable combination 
appearing in the FP.  In many cases, however, the magnitude and surface 
brightness uncertainties are dominated by zero-point uncertainties of
$0.03$--$0.05$ mag.
Table 3 presents the photometric data for the sample we consider.  
We include the uncertainties in the strongly correlated quantities 
($\mu_e-3.03\log \theta_e$) appearing in the FP equations.  
In our analysis we correct the photometric data for Galactic extinction 
using the extinction estimates
of Schlegel, Finkbeiner \& Davis (1998) and an $R_V=3.1$ extinction curve.

Of the \ntot$\,$ lenses for which we have photometry, we excluded \ndrop$\,$ from
the present survey.  B~0218+357, RXJ~0921+4528, B~1600+434 and PKS~1830--211 have  
late-type lens galaxies.\footnote{We also know that B~0218+357, B~1600+434 and PKS~1830--211 
are dusty (see Falco et al. 1999), and that B~0218+357 and PKS~1830--211 contain dense molecular gas 
(see Menten, Carilli \& Reid 1999 and references therein).}  Where the bulge of a 
late-type galaxy dominates the lensing, as in Q2237+0305, the bulge should lie 
on the fundamental plane and there is no reason for excluding the system.  The lens
galaxy remains undetected in Q~1208+1011 (Leh\'ar et al. 2000), and the
high contrast between the quasar images and the lens galaxy leads to poor
lens galaxy photometry in QJ~0158--4325 (CTQ414), SBS~0909+532, H~1413+117 and 
FBQ~1633+3134.  In MG~0751+2716 and B~1933+507 we are 
unable to reliably decompose the image into source and lens contributions. 
B~2114+022 (Jackson et al. 1998a) has two possible primary lens galaxies and 
the astrometric alignment of the lensed radio sources with the galaxies is still 
uncertain.  We expect 10-20\% of lenses to be late-type galaxies in the absence
of any population evolution (e.g. Fukugita \& Turner 1991, Maoz \& Rix 1993, 
Kochanek 1993, 1996).  Hence, the number of lenses we have dropped from the
analysis is consistent with no number evolution in the early-type population
to $z=1$.   

\subsection{Spectrophotometric Modeling}

We use the GISSEL96 version of the Bruzual \& Charlot (1993) spectral evolution
models to interpret the results.  The models were computed for $H_0=65\kms$ Mpc$^{-1}$
and three cosmological models (a flat cosmology with $\Omega_0=1$, a flat cosmology with 
$\Omega_0=0.3$ and $\lambda_0=0.7$, and an open cosmology with $\Omega_0=0.3$). We
initially considered three star formation histories:  an instantaneous burst in which all
stars form at redshift $z_f$, an extended burst in which star formation starts at
$z_f$ and continues at a constant rate for 1~Gyr, and an exponentially decaying star 
formation starting at redshift $z_f$ with a 1~Gyr e-folding time.  However, we found
that the interpretation was largely independent of the details of the star formation
history and we restricted our presentation to the instantaneous burst model because it 
makes the interpretation of $z_f$ straightforward.  The results for the extended or 
exponential burst models will match the instantaneous burst models if the end of the
star formation epoch (roughly $z_f$ plus 1~Gyr) is matched to the instantaneous 
burst redshift.  
In most cases we used a solar metallicity model ($Z_\odot=0.02$), although for 
redshift estimation (\S5) we used a range of models from $Z_\odot/4$ to $4Z_\odot$
to provide a range of galaxy colors at fixed age.

\subsection{Direct Comparison to Cluster Galaxies at Similar Redshifts}

The comparison to the spectrophotometric models is subject to systematic problems in
both the models and the data analysis.  By directly comparing the field, lens
population to the rich cluster population we can avoid these systematic problems.  
We obtained the archival WFPC2 
images of the rich clusters studied by van Dokkum \& Franx (1996), 
Kelson et al. (1997), van Dokkum et al. (1998a) and Pahre, Djorgovski \& de Carvalho (1999ab).  We extracted 
the early-type galaxies for which velocity dispersions were measured and 
redetermined their photometric properties using our standard methods (Leh\'ar et al. 2000)
to obtain a comparison sample of \nclst$\,$ early-type galaxies in very dense 
environments (see Tables 2 and 4).  Since the photometric reductions and zero-points are identical 
for the lens and cluster samples, any differences between the samples must be
caused either by differences in their star formation histories or by the methods
for estimating the central velocity dispersions of the lens galaxies (see \S6).  
Table 4 presents the data for the cluster galaxies.
For Cl1358+62 (Kelson et al. 1997), MS2053--04  
(Kelson et al. 1997) and MS1054--03 (van Dokkum et al. 1998a), where we are fitting the
same data, we agree with the published effective radius estimates to accuracies
of (mean$\pm$dispersion) $0.06\pm0.14$ dex, $-0.02\pm0.10$, and $-0.05\pm0.06$ dex 
respectively despite the different procedures used to determine the effective radius in 
each analysis.  Kelson et al. (2000) refit the Cl1358+62 galaxies, with a change in the 
effective radius of $-0.04\pm0.07$ relative to their previous results, and in closer 
agreement with our results ($0.02\pm0.09$).  The importance of these differences is 
further reduced by the insensitivity of the quantities appearing in the FP to 
errors in $r_e$.

\begin{figure}
\centerline{\psfig{figure=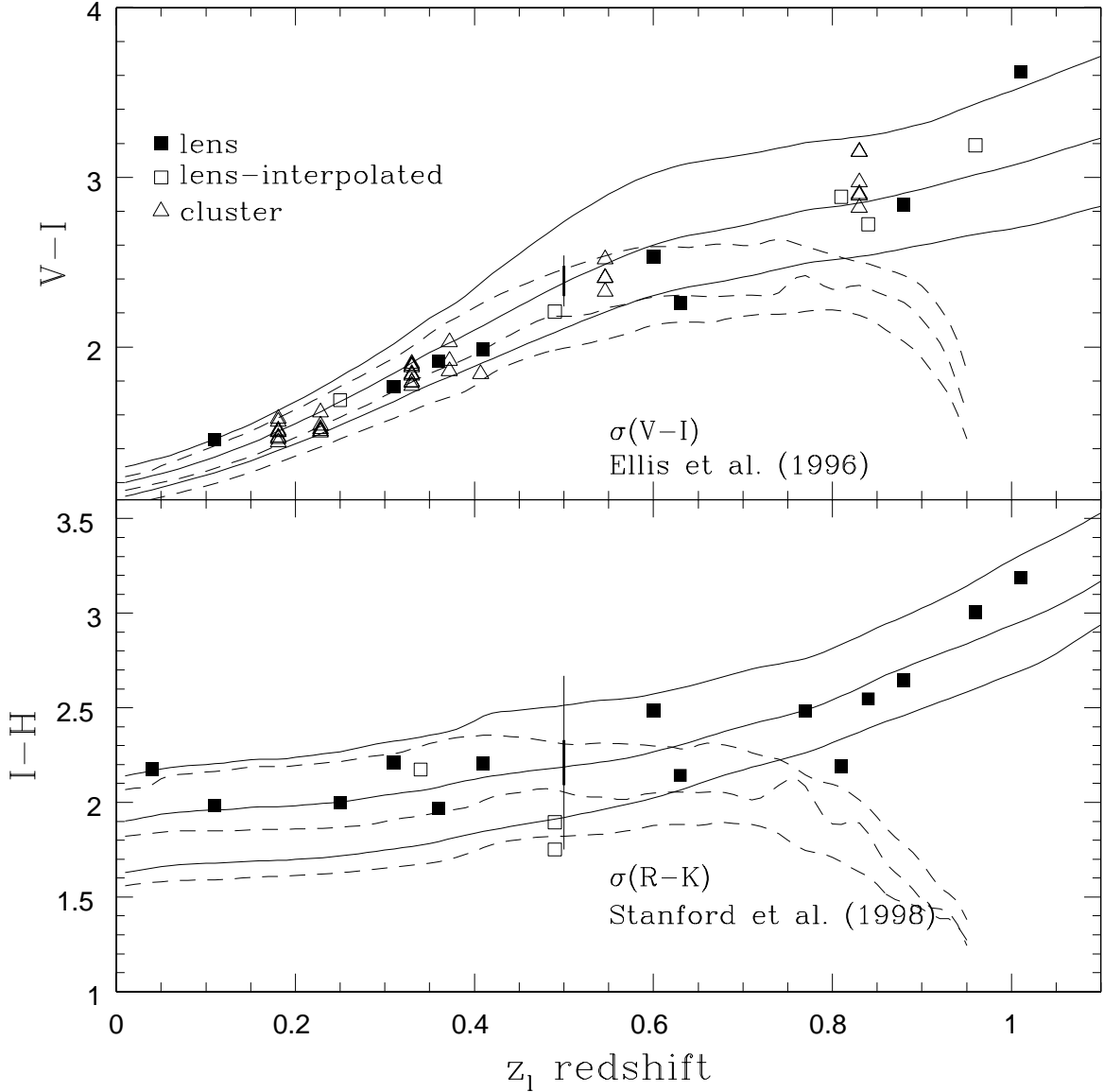,height=6.5in}}
\caption{V--I (top) and I--H (bottom) colors as a function of redshift for the lens and cluster galaxies.
  Only the lenses with spectroscopic redshifts are included.
  The heterogeneous optical filters for the lens data are interpolated to V=F555W or I=F814W as needed
  using the solar metallicity, $z_f=3$ instantaneous burst model.  The solid (dashed) lines show the 
  predicted V--I and I--H colors for the $z_f=3$ ($z_f=1$) instantaneous burst 
  spectrophotometric models with metallicities of $Z=0.4Z_\odot$ (bottom), 
  $Z_\odot$ (middle) and $2.5Z_\odot$ (top).  The error bars at $z=0.5$ show the scatter in the
  V--I and R--K colors 
  observed for rich clusters at $z\simeq0.5$ by Ellis et al. (1997) and Stanford et al. (1998).  The heavy 
  error bars are the scatter at fixed luminosity, and the light error bars are the
  additional scatter due to the correlations between luminosity and color in a sample
  averaged over a wide range of luminosities.  Only lenses with known lens redshifts are
  included.
   }
\end{figure}

\section{The Colors of Lens Galaxies }

Keeton et al. (1998) made the first systematic survey of the colors of
lens galaxies, finding that most were consistent with the predictions
for passively evolving early-type galaxies -- blue or star-forming
lenses are rare.  With the CASTLES photometry we can now examine both
the optical and infrared colors of the galaxies, although the
incomplete and inhomogeneous WFPC2 observations mean we must still use the
spectrophotometric models to transform the data for display.  Figure 1
illustrates the I--H and V--I colors of the individual lens and
cluster galaxies as compared to the instantaneous burst models.

At low redshift, both the lens and the cluster galaxies have colors
consistent with either the low ($z_f=1$) or the high ($z_f=3$)
formation epochs, particularly if there is freedom to adjust the
metallicity.  Above $z=0.5$ the models begin to diverge, and it
becomes trivial to distinguish between the two formation epochs by
$z=1.0$.  We will adopt the solar metallicity, $z_f=3$, instantaneous 
burst as our standard star formation history.
The models are a good global match to the colors of both the lens and
cluster galaxies.  If we estimate the difference between the observed
optical colors and the model colors for the same redshift, the mean
color difference and its dispersion relative to the models is
\clopt$\,$ mag for the V--I color of the cluster galaxies and
\lnopt$\,$ mag for the V--I and R--I color differences of the lens
galaxies.  The color dispersion includes that from the measurement
errors, which are generally larger for the lenses.  
For comparison, the color dispersion for the large samples of
rich cluster galaxies at $z\simeq 0.5$ studied by Ellis et al. (1997)
and Stanford et al. (1998) is $0.13$ mag in V--I.  We do not have
infrared magnitudes for the cluster galaxies, but the optical (V, R or
I) to infrared (H) color difference and its dispersion is \lnir$\,$
mag, compared to a dispersion of $0.29$ mag in the R--K color for the
large cluster samples (Stanford et al. 1998).  

\begin{figure}
\centerline{\hphantom{THIS OFFSET IS TO}\psfig{figure=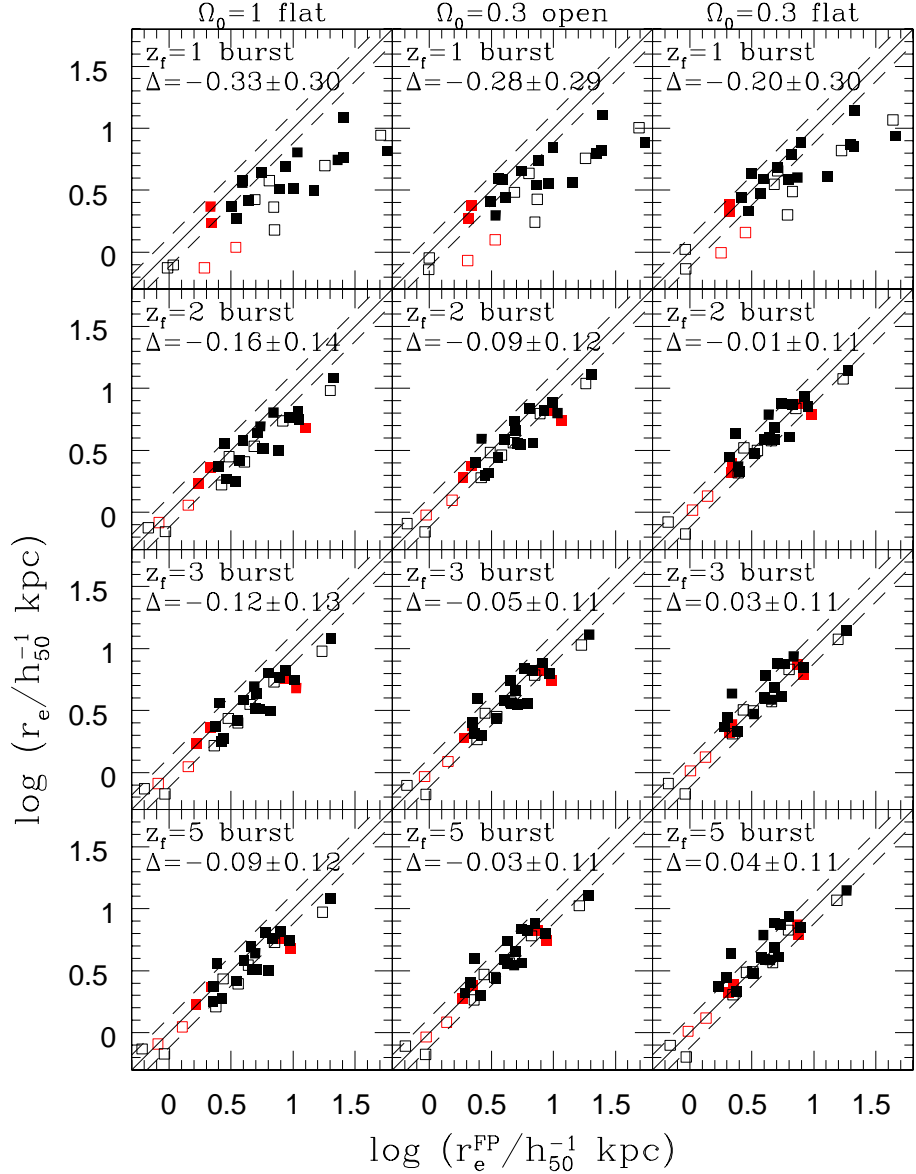,height=6.5in}}
\caption{The FP of lens galaxies transformed to zero redshift.  The cosmologies
  (from left to right) are the $\Omega_0=1.0$ flat, $\Omega_0=0.3$ open and
  $\Omega_0=0.3$ flat models.  The solar metallicity instantaneous burst star formation history 
  is used with star formation redshifts (from top to bottom) of $z_f=1$, $2$, 
  $3$ and $5$.  The filled squares are for the lenses with known redshifts and 
  for the open squares we have used the color and the FP to estimate the lens redshifts 
  (see \S5).  The solid line marks the FP of the local comparison sample; 
  90\% of the galaxies in the local JFK sample lie between the dashed lines. 
  The mean residual ($\Delta = \langle \log(r_e/r_e^{FP}\rangle)$) and its dispersion 
  are shown in the upper left corner of each panel.  These are calculated using only 
  the systems with known lens redshifts.  Lenses without spectroscopic redshifts are
  included using the methods of \S5 to estimate the lens redshift {\it assuming the 
  cosmological model and spectrophotometric model of that panel}.
   }
\end{figure}
\begin{figure}
\centerline{\hphantom{THIS OFFSET IS TO}\psfig{figure=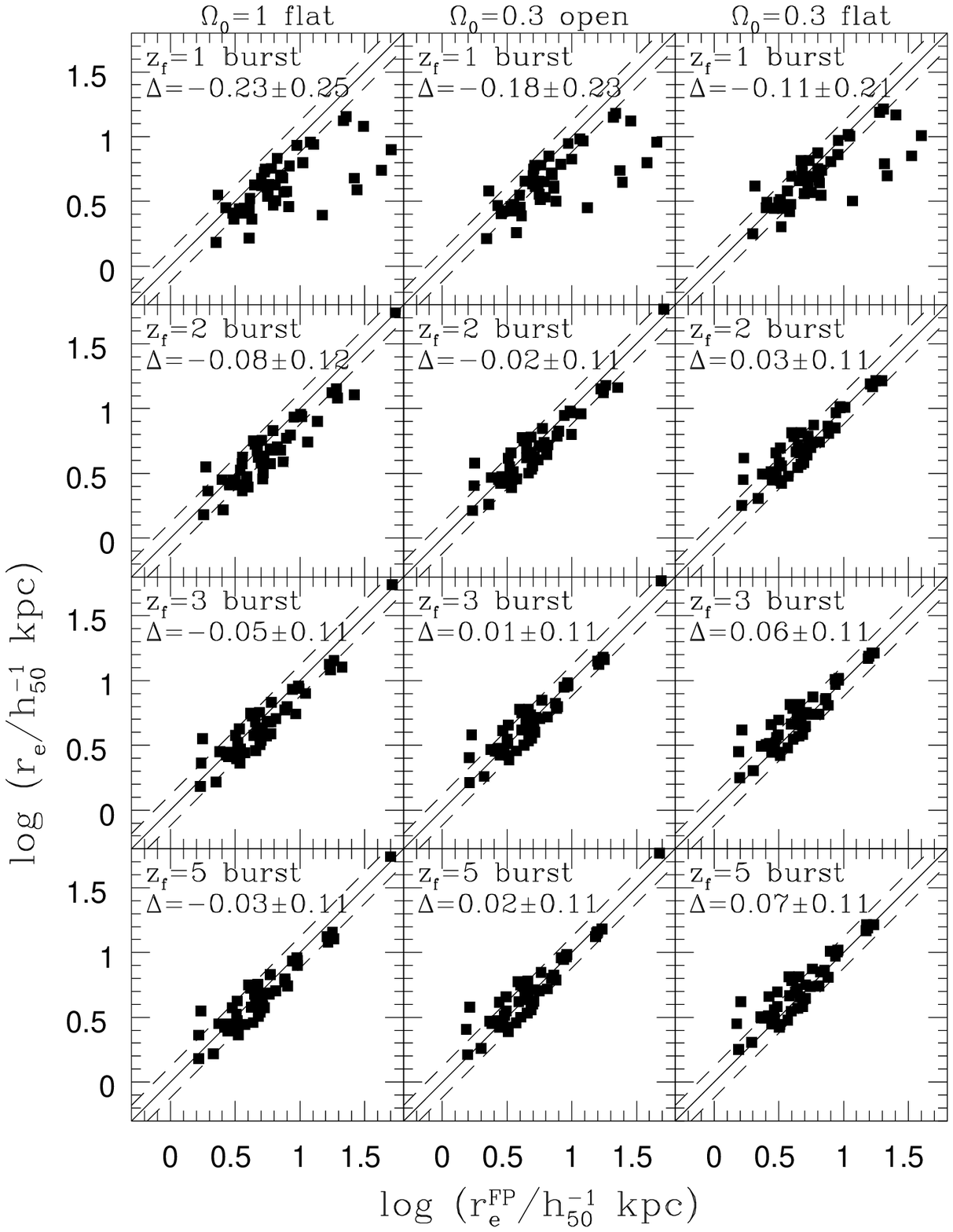,height=6.5in}}
\caption{The FP of cluster galaxies transformed to zero redshift.  Same cases
  as in Figure 2.  }
\end{figure}

\section{The Fundamental Plane of Lens Galaxies}

We can show that the lens galaxies lie on a coherent fundamental plane
by evolving their properties forward in time and placing them on the
present day FP.  The photometric transformations depend on both the
star formation epoch ($z_f$) and the cosmological model ($\Omega_0$
and $\lambda_0$), as illustrated in Figure 2 for the lens galaxies and
in Figure 3 for the cluster galaxies.  Where we possess data in
multiple filters, we have used the error-weighted average of the
estimates, and for the lens galaxies lacking spectroscopic redshifts
we have estimated the redshifts using the methods in \S5.  

It is clear from Figure 2 that the lens galaxies will lie on the local
FP if enough time has passed since the star formation epoch.  This
implies a star formation redshift $z_f \gtorder 2$ and a low matter
density cosmology.  As van Dokkum \& Franx (1996) noted for the
cluster early-type galaxies, it is difficult to reconcile an
$\Omega_0=1$ cosmology with a reasonable star formation epoch ($z_f
\ltorder 5$) unless significant changes are made in the initial mass
function (IMF) of the stars.  More remarkably, we see by comparing
Figures 2 and 3 that the FP of the lens galaxies is nearly identical
to that of the cluster galaxies.  The cluster galaxies show smaller
discrepancies for a low star formation redshift, but this 
is due to the lower mean redshift of the cluster sample.

We can quantify the differences using the mean offset of samples from the local FP,
$\Delta=\langle \log(r_e/r_e^{FP})\rangle$, and the dispersion of the sample around 
the mean offset,  $\sigma_\Delta$. For a population of galaxies
of the same age, we would find $\Delta=0$ and a minimum in $\sigma_\Delta$ when the
model age matched the true age. For a distribution of galaxy ages we would find $\Delta=0$
and a minimum in $\sigma_\Delta$ near the average age, but the width of the distribution
would be wider than that of the local FP.  In fact, the minimum dispersions of the two samples 
are identical, $\sigma_\Delta=0.11$, and larger than that of the local JFK sample
where $\sigma_\Delta=0.07$.  They are larger than the estimated uncertainties in the 
inputs to the calculation ($\log r_e -0.33\mu_e$ and $\Delta\theta$ or $\sigma_c$), 
which can be caused either by physics (the galaxy age distribution) or systematic errors 
(zero-points, averaging over heterogeneous filters, or the weak wavelength dependence of the FP).  
A physical explanation requires only a modest admixture of younger galaxies to a largely old
population.   For example, if fraction $\xi$ of the galaxy population formed at $z_f=1$ and fraction 
$1-\xi$ formed at $z_f=3$, we could explain the measured offsets and dispersions by an 
FP whose intrinsic width matches the local FP but has a $\xi=$13\% ($\xi=$35\%)
young galaxy fraction for the lens (cluster) sample in the $\Omega_0=0.3$ flat cosmology.
The lower mean galaxy redshift in the cluster sample allows a larger young galaxy fraction.  

The similar evolution of the cluster and lens galaxies is more reliably demonstrated by
the equivalence of the dispersions than by that of the offsets for the same assumed formation
epoch because the offset $\Delta$ is affected directly by systematic errors in estimating
the stellar velocity dispersion from the image separations using eqns. (5) or (6). 
In particular, we parameterized the uncertainty for the SIS model by the ratio 
$f=\sigma_D/\sigma_c$. Changing the value of this factor changes the offset $\Delta$
by $1.24\log f = \pm 0.12$ given the uncertainties estimated from stellar dynamics
and the distribution of lens separations ($f=1.0\pm0.1$).  Such freedom is sufficient
to significantly alter the formation redshift implied by the value of $\Delta$ up to 
a limit such that $z_f \gtorder 1.5$, but changing it has no effect on the dispersion
$\sigma_\Delta$ or the estimate of a possible young galaxy fraction $\xi$.
We discuss the dynamical normalization of the lenses further in \S6, and we 
measure the rate of galaxy evolution independent of the dynamical normalization
in \S7.
  
\begin{figure}
\centerline{
  \psfig{figure=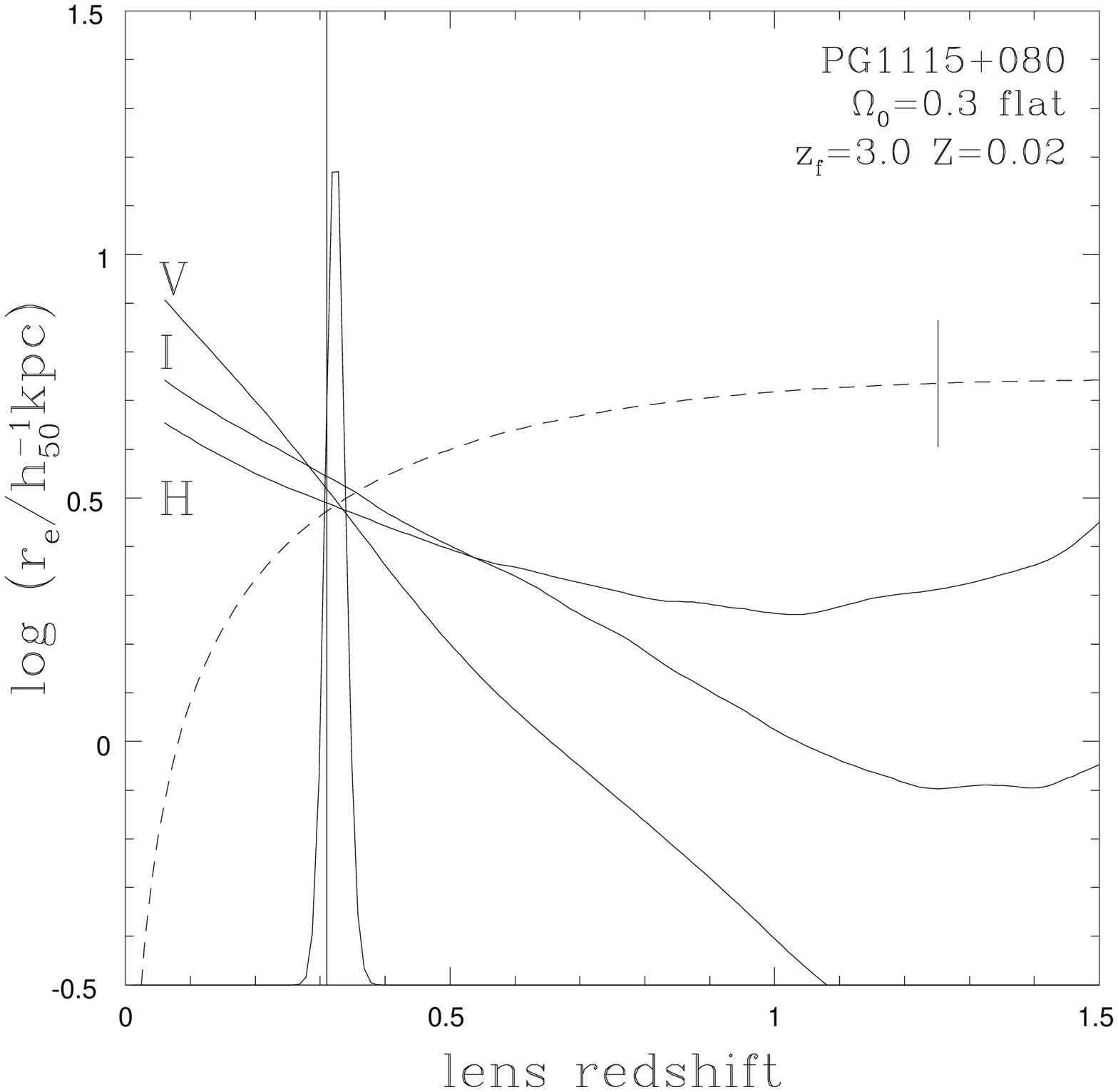,height=3.0in}\psfig{figure=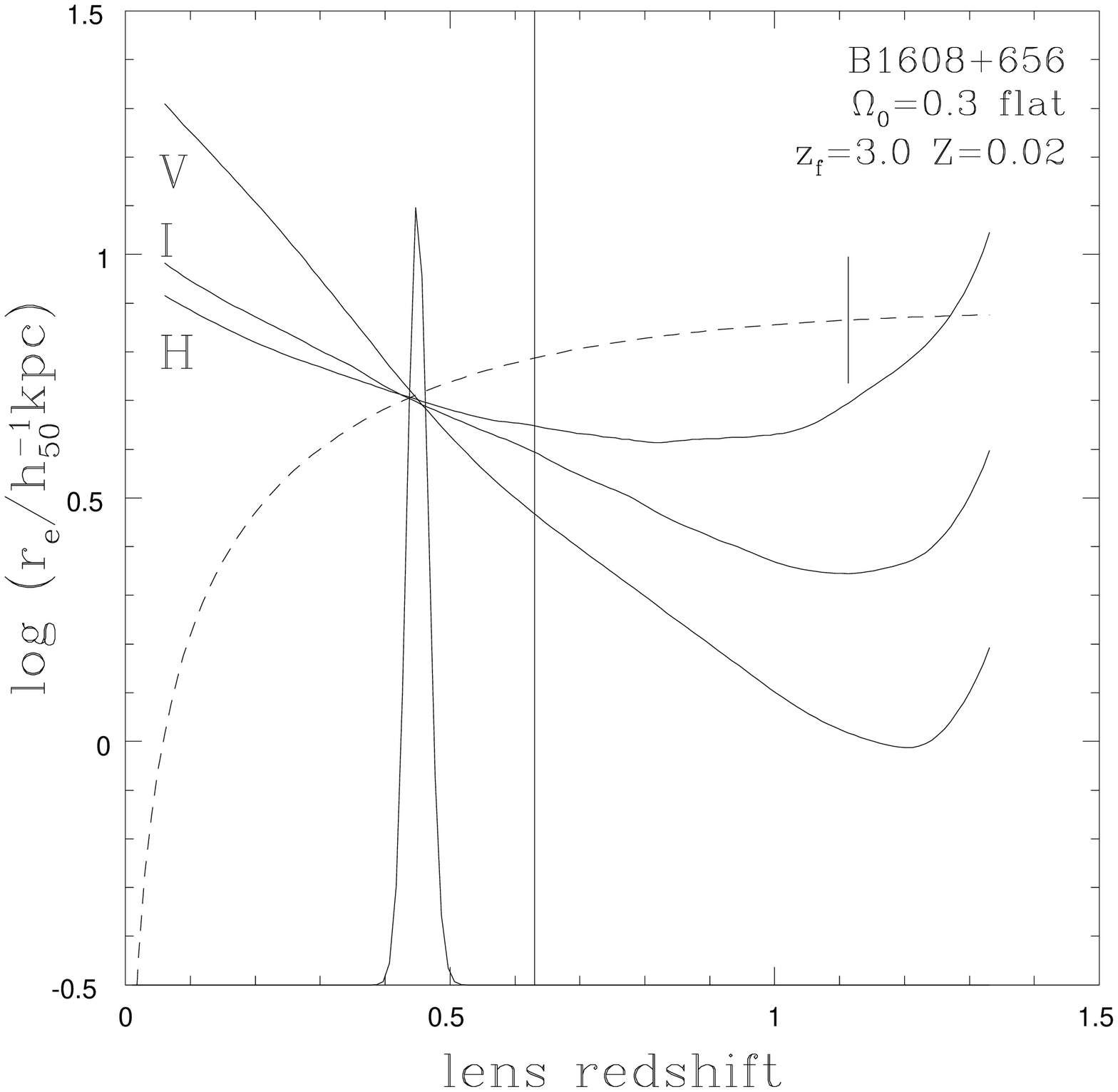,height=3.0in}}
\caption{ 
  Examples of redshift estimation for the $z_l=0.31$ lens PG~1115+080 (left) and
  the $z_l=0.63$ lens B~1608+656 (right).  The dashed curve is the physical effective
  radius $r_e$ as a function of redshift.
  The errorbar on the dashed curve would encompass 90\% of the galaxies 
  on the local FP.  The solid curves are the effective radii estimated using
  the FP ($r_e^{(j)}$) using the V, I and H surface photometry of the lens
  galaxy and assuming a solar metallicity, a $z_f=3$ instantaneous burst model 
  and an $\Omega_0=0.3$ flat cosmological model.  A vertical line marks the 
  spectroscopic lens redshift, and the ``Gaussian'' distribution shows the
  likelihood of fitting the data ($\propto \exp(-\chi^2/2)$).  A galaxy lies
  on the FP when a solid curve crosses the dashed curve, and the galaxy colors
  match the spectrophotometric model when all the solid curves intersect.
  }
\end{figure}
\begin{figure}
\centerline{\psfig{figure=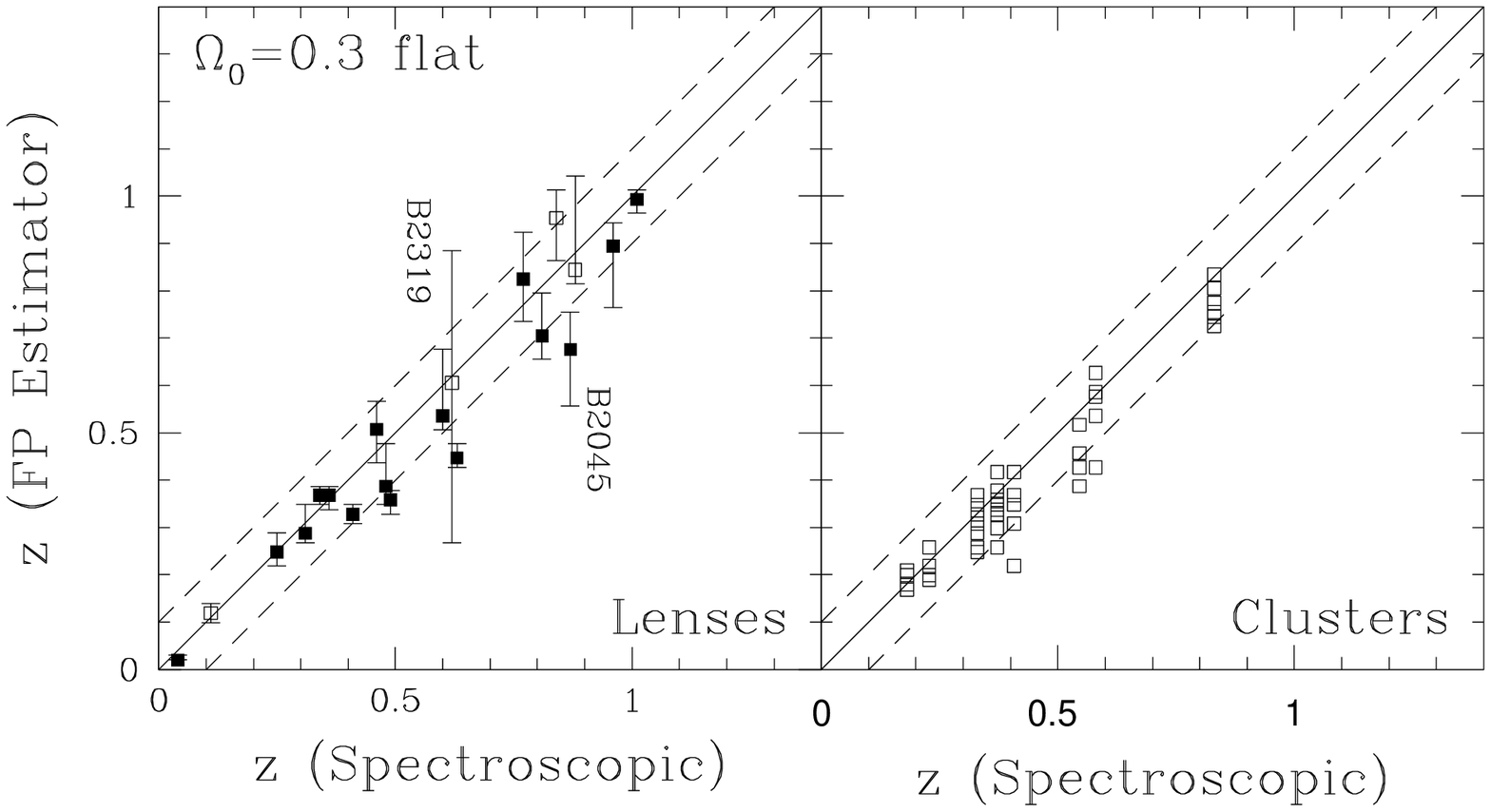,height=4.5in}}
\caption{The FP redshift estimates compared to the spectroscopic redshifts. 
  The left (right) panel shows the results for the lens (cluster) galaxies.
  For the lens galaxies, solid (open) points are used for 
  lenses where the source redshift is known (unknown).  The dashed lines are
  offset by $|\Delta z|=0.1$ to illustrate the desired accuracy.  We have only
  H-band data for the labeled lenses with large redshift uncertainties 
  (B~2045+265 and B~2319+052). }
\end{figure}

\section{Redshift Estimation}

Unmeasured lens redshifts are a major limitation on the use of gravitational
lenses to study cosmology (see Kochanek 1992, Helbig \& Kayser 1996, Kochanek 1996),
so it is important to develop a reliable, accurate method for estimating lens
galaxy redshifts.  The redshift uncertainties from estimates based on the Faber-Jackson 
(1976) relation or gravitational lens statistics are too broad (see Kochanek 1992),
even though Keeton et al. (1998) found that lens galaxy luminosities follow the expected
correlations.  The reduced scatter in the FP compared to the Faber-Jackson
relationship should make it a better means of estimating lens redshifts.
Moreover, we now have accurate colors for many of the lens galaxies, and
photometric redshifts should work very well because most lens galaxies are
intrinsically red, early-type galaxies.

We estimate the redshifts using the formalism of the fundamental plane
because it leads to a method that works in the absence of any color
information.  For each filter $j$ we use a spectrophotometric model
and the FP relations (eqn. 1) to estimate the physical effective
radius, $r_e^{(j)}(z_l,z_s)$, and then compare it to the observed
physical effective radius, $r_e$.  The
estimates of the effective radius derived from the FP ($r_e^{(j)}$)
depend on the lens redshift, the source redshift and the
spectrophotometric model, while the estimate of the physical effective
radius ($r_e$) depends only on the lens redshift through the angular
diameter distance.  When we match all the estimates $r_e^{(j)}$ we are
obtaining a color redshift, and when we match $r_e^{(j)}$ with $r_e$
we are obtaining a fundamental plane redshift.  A $\chi^2$ statistic
measures the goodness of fit given the magnitude uncertainties and
the thickness of the fundamental plane. 

Figure 4 illustrates the method for two systems and a particular
model (a $z_f=3.0$ solar metallicity instantaneous burst).  A perfect
example of the method is PG~1115+080, where the galaxy lies precisely 
on the FP with colors matching the spectrophotometric model at the true 
lens redshift.  An example of a self-consistent, but incorrect, redshift
determination is shown by B~1608+656.  As in PG~1115+080,
there is a redshift where the galaxy lies precisely on the FP with
colors matching the model, but it is significantly offset from the
true lens redshift and the estimated uncertainty is far less than the
actual error.  The B~1608+656 lens galaxy has both $[\hbox{O~II}]$
emission and a weak $4000\AA$ break (Myers et al. 1995), indicating
some ongoing star formation.  If we use a $z_f=1$ spectrophotometric
model instead of a $z_f=3$ model, then the lens both lies on the FP
and matches the model colors at its true redshift.  There are few
such lens galaxies, consistent with the low young galaxy fraction
we estimated in the previous section.  For both of these systems
we would have obtained the same redshift estimate from any single
filter (no color information) just from the requirement that the
galaxy lie on the FP.  Such single filter redshift estimates work
best at low redshift, where the angular diameter distance changes
rapidly, or with optical filters, where the K-correction changes
rapidly.  The accuracy will be poor given only an infrared 
observation of a lens galaxy at $z_l \gtorder0.5$ where both   
the distance and the surface brightness vary slowly with redshift.

We determined the final redshift estimates by fitting each lens using $z_f=3$ 
instantaneous burst models with a range of metallicities ($0.4Z_\odot$ to $2.5Z_\odot$)
in an $\Omega_0=0.3$ flat cosmological model.  The results are insensitive
to the cosmological model, using a range of metallicities (a range of colors at
fixed age) significantly improved the accuracy, and using a range of star formation 
epochs ($z_f=1$ to $z_f=3$) decreased the accuracy.  For each lens we found
the model and lens redshift which best fit the data and then estimated the
uncertainties by the redshift range such that the change in the $\chi^2$
statistic satisfied $\Delta\chi^2<4$ including variations in the metallicity. 
The results are illustrated in Figure 5.  Formally, the uncertainty estimates
are $2\sigma$ limits, but many of the errors are systematic rather than statistical
and the $\Delta\chi^2<4$ error estimate seems to match the actual scatter seen
in Figure 5. The mean and dispersion of the redshift differences are 
$\langle z_{FP}-z_l \rangle = \lnzall$ for the lenses and 
$\langle z_{FP}-z_c \rangle = \clzall$ for the clusters.  The redshift estimates 
are presented in Table 5.  Note that the redshifts derived in Figure 4 using a particular 
spectrophotometric model will differ from the final estimates in Figure 5 and Table 5 
which statistically average over a range of models.
  
\def\sigthet{\sigma_{\Delta\theta}}
\begin{figure}
\centerline{\psfig{figure=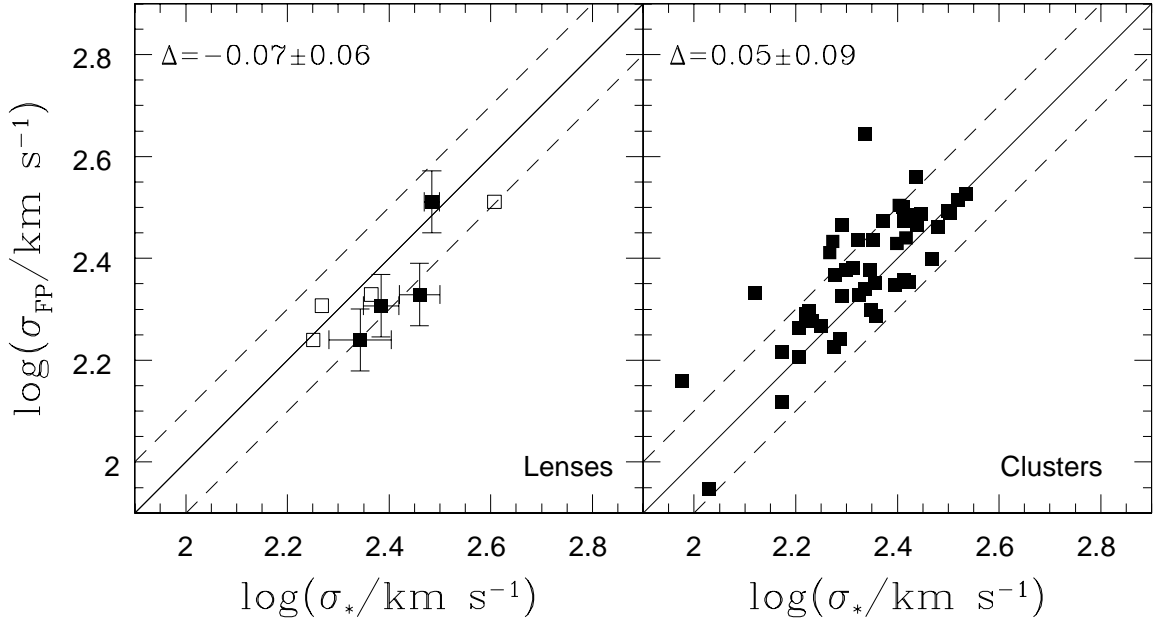,height=4.5in}}
\caption{A comparison of central velocity dispersions, $\sigma_*$, and velocity dispersion 
  predictions from the FP, $\sigma_{FP}$, for the lens galaxies (left) and the cluster galaxies (right).  
  The lens galaxies are Q2237+0305, MG1549+3047, PG1115+080 and Q0957+561 (in order of increasing $\sigma_*$).  
  For the lenses, the velocity dispersions estimated from the image separations, $\sigthet$, are shown as open points.
  The dashed lines would encompass 90\% of the galaxies in the local JFK sample.  The mean offset 
  $\Delta=\langle \log\sigma_{FP}-\log\sigma_*\rangle$ and its dispersion are shown in the upper left
  corner.  The value of $\sigma_{FP}$ was computed for the solar metallicity, $z_f=3$ instantaneous
  burst model and an $\Omega_0=0.3$ flat cosmological model.  Note that the right panel is the same
  as the equivalent model in Figure 3.   
  }
\end{figure}
\begin{figure}
\centerline{\psfig{figure=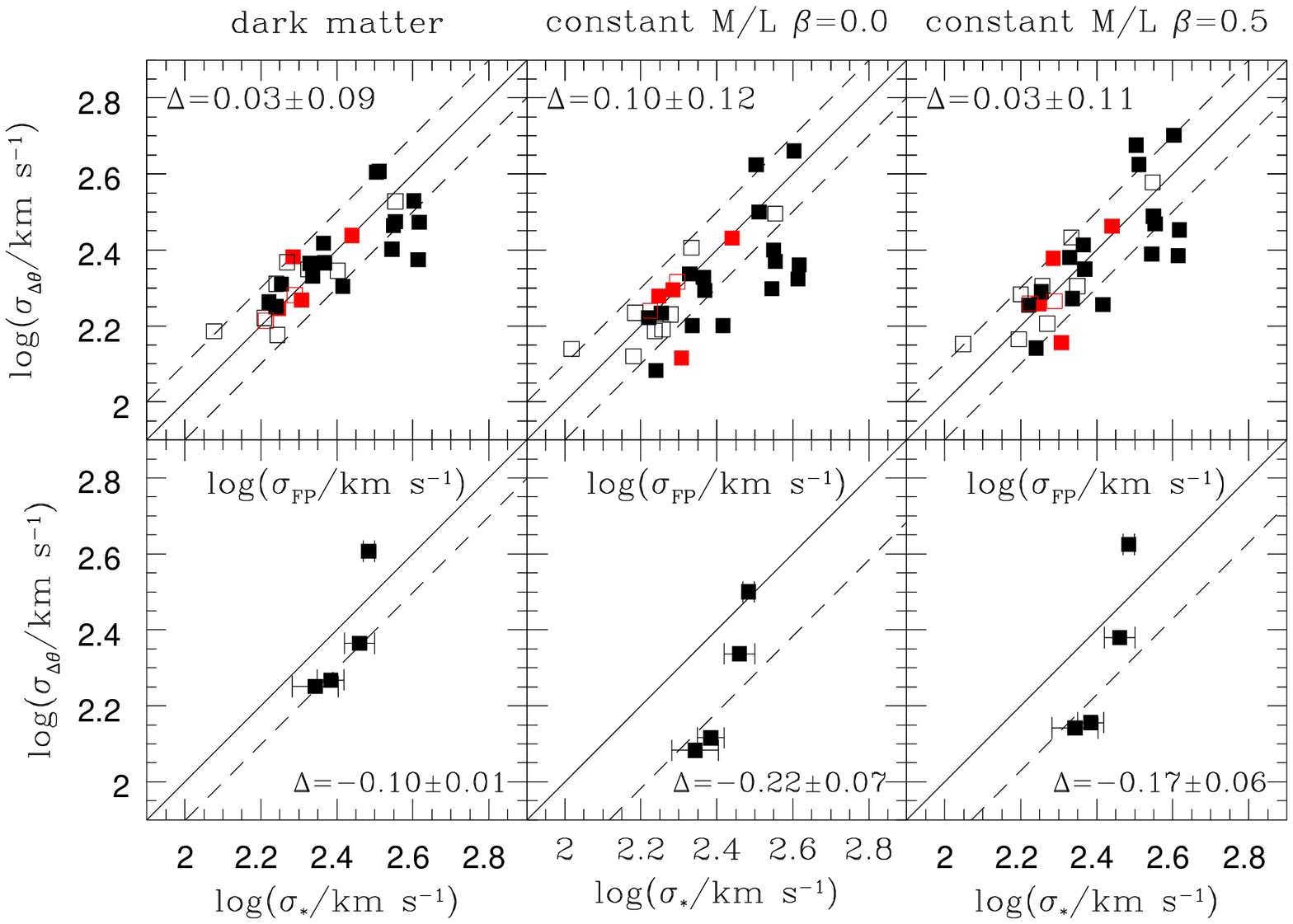,height=4.5in}}
\caption{A comparison of the velocity dispersions estimated from the lens model, $\sigma_{\Delta\theta}$, to 
  the measured stellar dispersion, $\sigma_*$ (bottom), and to the dispersion estimated from the FP, $\sigma_{FP}$
  (top).  The value of $\sigma_{\Delta\theta}$ is estimated from the lens geometry using the dark matter (left), 
  isotropic constant $M/L$ (middle), and radially anisotropic constant $M/L$ (right) dynamical models.  In the top 
  panels, $\Delta=\langle\sigma_{FP}-\sigthet\rangle$ to match the form used in Figures 2 and 3.  In the  
  bottom panels, $\Delta=\langle\sigthet-\sigma_*\rangle$, but we exclude Q~0957+561 because of the effects 
  of the cluster on $\sigthet$.  The values are the mean and the dispersion about the mean.
  The cosmological and spectrophotometric models are the 
  same as in Figure 6.  In the upper panel the dashed curves would encompass 90\% of the galaxies in the local 
  JFK sample, and in the lower panel the dashed curve is the mean offset excluding Q~0957+561.  
  }   
\end{figure}

\section{Dynamical Normalization}

We next consider the dynamical normalization of the lens models using the FP as a tool to probe
the relationship between $\sigma_*$ and $\Delta\theta$.   This is a substantial digression from
the question of galaxy evolution, and in \S7 we will rederive the evolution of the galaxies
in a way which is clearly independent of uncertainties in the normalization.  Nonetheless, consideration
of gravitational lenses and the FP would be incomplete without further discussion of 
normalization.  Ideally we would like to have a large sample of gravitational lenses with direct
velocity dispersion measurements. In fact, velocity dispersions have been measured for four 
gravitational lenses Q0957+561 (Tonry \& Franx 1999, Falco et al. 1997, Rhee 1991), PG1115+0808 
(Tonry 1998), MG1549+3047 (Leh\'ar et al. 1996) and the bulge of Q2237+0305 (Foltz et al. 1992).  
Using the aperture correction model of Jorgensen et al. (1995) we estimated that the velocity dispersions of
the four lens systems in the standard $3\parcs4$ aperture at Coma are $\log(\sigma_c/\kms)=2.484 \pm 0.015$, 
$2.460\pm0.040$, $2.384\pm0.035$, and $2.343\pm0.061$ respectively.  The errors include the formal measurement
error combined in quadrature with a generous uncertainty in the aperture definition.  Unfortunately,
two of these four systems are peculiar: Q0957+561 is the lens whose properties are most contaminated
by the mass distribution of the cluster to which the lens galaxy belongs, and Q2237+0305 is a bright
nearby spiral galaxy rather than a massive early-type galaxy.  Nonetheless, we have three estimates  
of the velocity dispersion to compare: the directly measured stellar dispersions, $\sigma_*$; the 
estimates of the stellar dispersion from the image separation, $\sigthet$, which will depend on the 
dynamical model and (weakly) on the cosmological model; and the FP estimate of the velocity dispersion, 
$\sigma_{FP}$, which depends on the cosmology and the spectrophotometric model.  

We first compare the measured stellar dispersions to the FP predictions ($\sigma_*$ and $\sigma_{FP}$) because
this comparison is independent of the dynamical model used for the lenses.  Figure 6 shows the comparison
for both the lens and cluster galaxies assuming the solar metallicity, $z_f=3$ instantaneous burst model.
In the local JFK sample the dispersion of the FP measured in terms of velocity dispersions is $0.06$ dex with 
90\% of the galaxies lying within 0.1 dex.  For the clusters, Figure 6 is simply the comparison made
in Figure 3 plotted in terms of velocity dispersion rather than effective radius.  
For the lenses this comparison is fundamentally different from that in \S3 and Figure 2  
because we are not using the lens geometry to 
estimate the velocity dispersion.  If we estimate the offset by $\Delta = \langle \sigma_{FP}-\sigma_*\rangle$, 
then the velocity offset should be $0.8$ of the effective radius offset measured in \S3 because it is just
a rearrangement of the terms in the FP defined by eqn. (1).  To the extent that four points suffice
to define the FP, we see that the lenses will also lie on the FP if we use the directly measured velocity
dispersions, but that it is offset slightly from the estimate using $\sigthet$ in \S3. 

For the lenses we can also compare the velocity dispersion estimated from the image separation, $\sigthet$,
to the observed stellar dispersions, $\sigma_*$, and the FP estimates, $\sigma_{FP}$, as shown in Figure 7.
Here the results depend on the dynamical model, and we show the comparison for our standard dark matter model 
and the two constant M/L models (isotropic $\beta=0$ and radially anisotropic $\beta=0.5$).  The dispersion
in the FP (Figure 7, top) is roughly independent of the dynamical model, although the isotropic constant
M/L model has a significant offset from the FP found using either the dark matter or radially anisotropic
constant M/L model.  The sense of the offset is that the isotropic model would require an older stellar
population to match the local FP.  The agreement between $\sigthet$ and the measured stellar dispersions
$\sigma_*$ is significantly worse in the constant M/L models than in the dark matter model.  In this
comparison we must ignore Q~0957+561 where we know that the value of $\sigthet$ is increased by the effects of
the cluster containing the lens galaxy.  

Although the offsets are smallest for the dark matter model, they still imply a marginally significant 
normalization change to $f\simeq0.80\pm0.05$ rather than $f=1.0\pm0.1$ (see \S2.3).  Such a change in
the normalization requires either making the lens galaxies older or giving them more
dark matter on the scale of the image separations than for the galaxies in rich clusters. 
It is, however, a sample of only 3 objects, one of which is the bulge of a nearby spiral rather than
a massive early type galaxy.  The low value of $f$ also implies a separation distribution for the
lenses which would be grossly inconsistent with the data.\footnote{The average image separation 
scales as $\langle\Delta\theta\rangle \propto f^2$, so a change from $f=1$ to $f=0.8$ reduces
the mean image separation to 64\% of its standard value.}  Three of the four systems (Q~0957+561,
PG~1115+080 and Q~2237+0305) have bright quasar images in the core of the galaxy whose 
emission complicates velocity dispersion measurements, and it is possible that the significance of the
offset is exaggerated by underestimation of the uncertainties in the dynamical measurements. 
In \S8 we provide a list of lenses with relatively high central surface brightnesses and minimal source
contamination which could be used to improve the dynamical comparison.    
It would be best, however, to eliminate completely the effects of the normalization from the measurement 
of the evolution.    

\section{Galaxy Evolution}

We can eliminate the apparent dependence of the evolutionary estimates on the dynamical normalization
by abandoning the construction of the FP (as in \S4) and instead calculate directly the E+K corrections as
a function of lens redshift using eqn. (4).  The information on galaxy evolution is contained in the
changes in the corrections with redshift, and is independent of errors in the dynamical normalization.
In doing so we only assume that the galaxies will evolve 
to lie on the local FP at $z=0$.  The resulting values for the E+K corrections are independent of the 
spectrophotometric model but depend on cosmology through the variations in cosmological distances.
Note that the calculation of the evolutionary corrections only requires the existence of a coherent, thin
FP for the local reference population.  
To condense the results we transformed the non-standard filters into the standard 
filters using the best fit spectrophotometric model.\footnote{We transformed J=F110W and K=F205W to H=F160W, 
F702W and F791W to I=F814W, and R=F675W and F606W to V=F555W.}
Where the filter transformations lead to multiple estimates for a single 
lens, we decided to regard it as an additional means of estimating uncertainties.  The surface
brightness scatter of the FP is still significant (in the local JFK sample it 
is $0.23$ mag), so we averaged the estimates from the individual galaxies in four redshift bins 
with edges at $z=0.25$, $0.50$ and $0.75$ to produce the final numerical
estimates presented in Table 6.   The estimates of the E+K corrections depend on the spectrophotometric
models only through the filter transformations and lens redshift estimates.
  
The evolution with redshift is shown in Figures 8 and 9, from which we can immediately draw
five qualitative conclusions.  First, the FP makes the evolution of the galaxies obvious.  
With the $(1+z)^4$ cosmological dimming removed, the galaxies become rapidly fainter in the
V band, have almost constant surface brightness in the I band, and become steadily brighter
in the H band as the dominant term switches from the strong K-corrections at V band to the 
evolution corrections at H band.  Second, the evolution is positive in all three bands, with 
the surface brightnesses steadily rising above the predictions for a non-evolving population.  
Third, as van Dokkum \& Franx (1996) originally noted for the cluster galaxies, it is difficult 
to reconcile the measurements with the high $\Omega_0=1$ model because they 
require unphysically high star formation redshifts, $z_f > 10$.  Fourth, neither sample
is easily reconciled with low star formation redshifts ($z_f \ltorder 2$).  Fifth, there
are no obvious differences in the evolutionary histories of the early-type galaxies in
low and high density environments.  
  
Dynamical normalization errors have no effect on these evolution estimates, because they
depend only on the changes in the E+K corrections with redshift.  A change 
in the normalization factor $f$ for the lenses shifts the estimates by $\Delta(e+k)=-3.76\Delta\log f$
for all filters and at all redshifts -- changing $f$ changes the zero-redshift mass-to-light 
ratio of the galaxies but not its evolution.  If we fit the E+K corrections with a 
linear function of redshift, extrapolate to zero-redshift and estimate the difference between
the estimates for the lens and cluster galaxies, we find $\Delta(e+k)=0.05\pm0.25$ and $-0.14\pm0.24$ for 
the V and I-bands respectively.  The individual estimates of the $E+K$ corrections are
all statistically consistent with zero at redshift zero.
We know from the redshift evolution that the two populations
have similar star formation histories, so the zero-redshift difference in the E+K corrections
should provide an accurate means of estimating the normalization factor.  Averaging the V
and I-bands we obtain a final estimate of $f= 1.06\pm0.07$, which is consistent with our
original estimates in \S2 from either stellar dynamics or the average image separations of the
lenses. 

\begin{figure}
\centerline{\psfig{figure=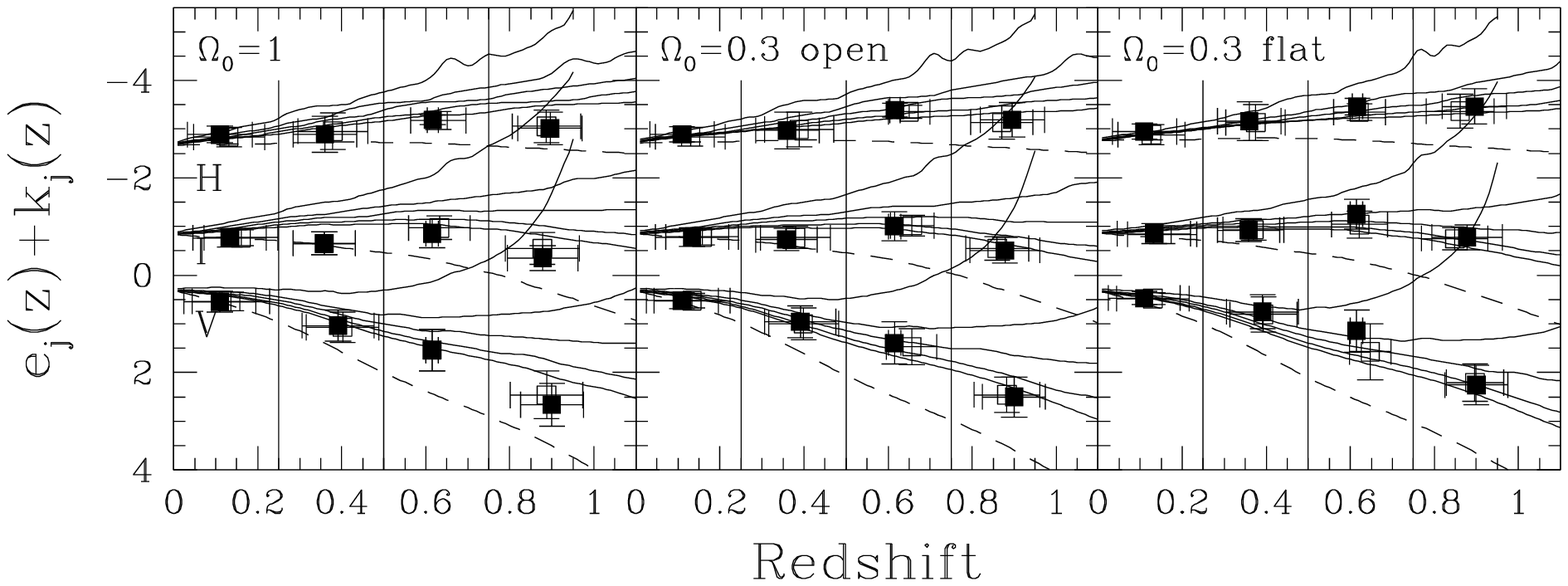,width=6.5in}}
\caption{Evolution and K-corrections for the lens sample in the
V=F555W, I=F814W and H=F160W bands as a function of redshift for the
$\Omega_0=1$ (left), $\Omega_0=0.3$ open (middle) and $\Omega_0=0.3$
flat (right) cosmological models. The zero-redshift color differences
were left in to separate the curves.  
The results are averages for all
galaxies within the bins delineated by the vertical lines.  
The points are located at the mean redshift, and
the redshift errorbar is the standard deviation of the redshift
distribution in each bin.  The error in the E+K correction is the
standard deviation of the points, not the uncertainty in the mean
(which would be smaller by $1/(N_{bin}-1)^{1/2}$ where $N_{bin}$ is
the number of points in each bin).  The filled points are the averages
using only the lenses with known lens redshifts, while the open points
include all lenses.  The dashed curves are the no evolution models for
each filter, and the solid curves are the instantaneous burst models
with star formation redshifts (from bottom to top) of $z_f=10$, $3$,
$2$, $1.5$ and $1.0$ respectively.  }   
\centerline{\psfig{figure=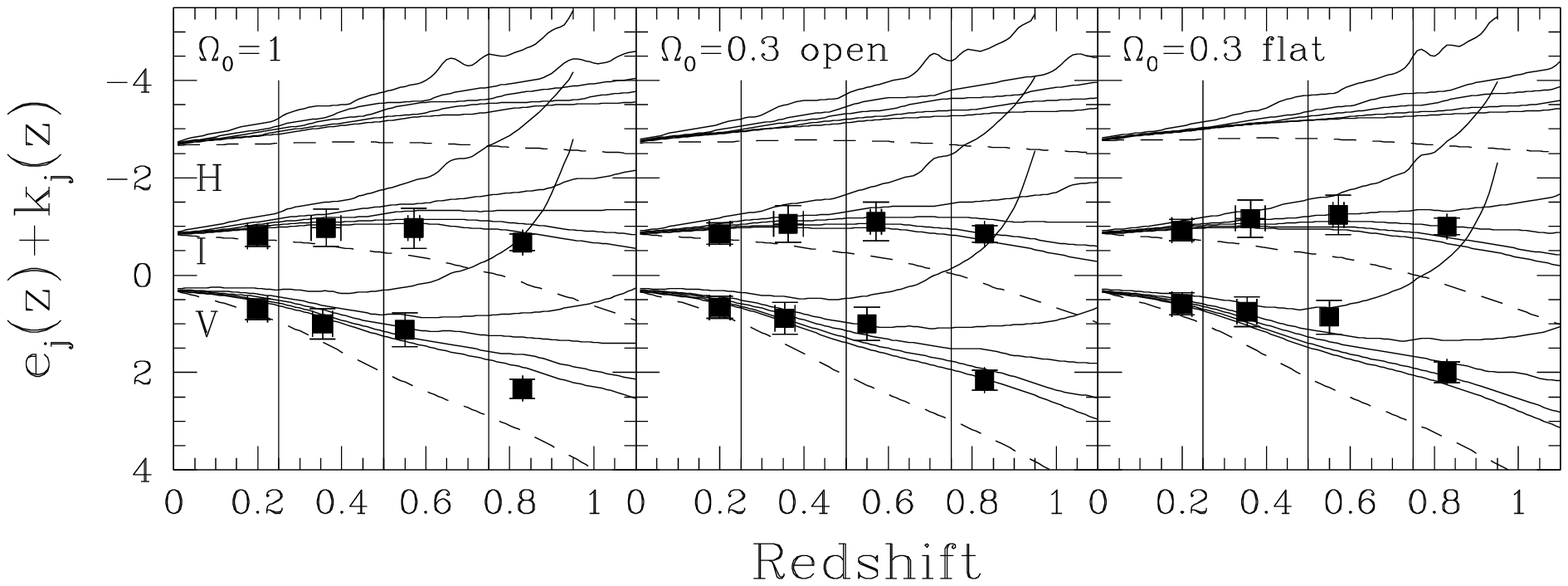,width=6.5in}}
\caption{Evolution and K-corrections for the cluster sample in the V=F555W and I=F814W bands
  as a function of redshift for the $\Omega_0=1$ (left), $\Omega_0=0.3$ open (middle) and $\Omega_0=0.3$ flat
  (right) cosmological models.  The format is identical to that of Figure 8.  
  }
\end{figure}

\section{Conclusions}

Most gravitational lens galaxies are early-type galaxies lying on the passively
evolving fundamental plane, allowing us to measure the evolution rate of early-type 
galaxies in low density environments.  We find that the stars constituting the lens
galaxies must have formed at $z_f \gtorder 2$ for an $\Omega_0=0.3$ flat cosmology.  
The required formation epoch increases for an open or higher matter density 
cosmology.  Star formation histories with an extended burst or an exponentially decaying burst
must form most of their stars before this redshift limit.  More generally, we
have directly measured the E+K corrections for ``field'' early-type galaxies
in the V, I and H bands over the range $0< z < 1$.  

When we compare the lens galaxies to those in FP studies of rich clusters 
(van Dokkum \& Franx 1996, Kelson et al. 1997, van Dokkum et al. 1998a, 
Pahre, Djorgovski \& de Carvalho 1999ab, 
Jorgensen et al.  1999), we find no significant differences between 
the galaxies in the two environments.  The high redshift of the star formation epoch and the
lack of a difference between the two environments are not
consistent with the predictions of standard semi-analytic models of galaxy 
formation (Kauffmann 1996, Kauffmann \& Charlot 1998), where the stellar
populations of the early-type galaxies in low-density environments (groups
and clusters with velocity dispersions $<400\kms$) are 
predicted to have formed at $0.5 \ltorder z_f \ltorder 1.5$.
Somerville \& Primack (1998) have argued more generally that the earlier
semi-analytic models (Kauffmann 1996, Kauffmann \& Charlot 1998, 
Baugh et al. 1998) have systematically underestimated the typical star
formation epoch through their choice of star formation mechanisms.  
Kauffmann \& Charlot (1998) would argue, however, that early-type galaxies look
old at all redshifts because the bulk of the star formation took place when the
galaxies were morphologically late-type galaxies and that early-type galaxies are not
assembled until the stellar populations look old.  Such a bias is certainly
present in the cluster samples, where many of the galaxies are morphologically selected,
and galaxies with signs of recent star formation (E+A galaxies) are 
excluded from some of the samples (see van Dokkum \& Franx 1996, Kelson et al. 1997, 
van Dokkum et al. 1998a, Pahre et al. 1999ab, Jorgensen et al. 1999).  
Such galaxies are a small fraction of their samples
(10\% overall, 20\% in the highest redshift cluster), and so represent a modest
bias.  This argument also requires rapid number evolution in the early-type 
galaxy population by $z=1$, for which there is little evidence in the field 
(e.g. Schade et al. 1999). Moreover, the lens systems were selected 
based on their mass, not their morphology.  Studies of the effects of merging on 
gravitational lenses (Mao 1991, Mao \& Kochanek
1994, Rix et al. 1994) concluded that the total number of lenses would be preserved
but the balance would steadily shift to smaller image separations and late-type lenses
at higher redshifts if there was rapid number evolution in the early-type galaxies.  
While we omitted \ndrop$\,$ lenses of the \ntot$\,$ for which
we had obtained data, only 4 were dropped because they were clearly late-type 
lenses.  In the absence of any population evolution we would have 
expected 4 to 8 late-type lenses in the sample.
The others were dropped for the quality of the data or for possessing 
multiple lens galaxies.  None of these systems have the blue, high surface
brightness stellar populations which we would expect from the Kauffmann \& Charlot
(1998) scenario for how morphologically selected early-type galaxies could always 
appear to be old.  

Missing lens redshifts are a major barrier to using the lenses as 
astrophysical and cosmological tools, so we also explored using photometric redshift
estimates for the lens galaxies.  Galaxy colors, particularly with the
limited galaxy type range of the lenses, produce very good redshift estimates.
The FP is also a good redshift estimator, but redshift estimates from the FP
can have broad degeneracies when using infrared data for high redshift lens
galaxies. Our redshift estimation method has an empirical accuracy of 
$\langle z_{FP} - z_l \rangle = \lnzall$ for the \nlenz$\,$ lenses with 
known redshifts.  The scatter is dominated by lenses for which
we possess only H-band data, and would be reduced by the inclusion of
accurate infrared to optical color measurements.
Small amounts of late-time star formation reduce the accuracy of our estimates,
but lens galaxies with significant star formation or younger stellar populations
will usually have easily measured spectroscopic redshifts. 

In most systems we can only estimate the velocity dispersion of the lens galaxy from the
geometry of the lensed images.  Comparisons of these estimates both to the velocity
dispersions predicted from the fundamental plane and direct measurements for
four of the lenses indicated that a dark matter-dominated model is more consistent
with the data than constant M/L models, and that a radially anisotropic constant
M/L model was significantly better than an isotropic model.  Our results for the
normalization factor $f=\sigma_D/\sigma_c$ of the dark matter model are somewhat ambiguous, 
as we find $f\simeq0.80\pm0.05$ if we match the three usable lens galaxy velocity 
dispersions and $f=1.06\pm0.07$ if we match the zero-redshift mass-to-light ratios of 
the lens and cluster galaxies.  The central surface brightnesses of the 
lens galaxies are similar to those of the cluster galaxies, so the velocity dispersions 
can usually measured if there are no nearby, bright source images.  Five good candidates with surface 
brightnesses in a 1\farcs0 aperture above 23 R mag/arcsec$^2$ are MG1654+1346 
($\langle\mu_R\rangle=20.2$), 0047--2808 ($\langle\mu_R\rangle=21.1$), 
B0712+472 ($\langle\mu_R\rangle=21.3$), B1608+656 
($\langle\mu_R\rangle=21.7$), HST1411+5211 ($\langle\mu_R\rangle=22.2$) and 
HST14176+5226 ($\langle\mu_R\rangle=22.8$).  The systems with measured velocity
dispersions had central surface brightnesses of $17.7$, $19.1$, $20.2$ and $21.1$~R~mag/arcsec$^2$
for Q~2237+0305, MG~1549+3047, Q~0957+561 and PG~1115+080 respectively.  We would not
have included Q~2237+0305, Q~0957+561 and PG~1115+080 in our list of dynamical targets
because they also have bright quasar images which will contaminate the galaxy spectrum.
With a larger number of directly measured dispersions, more sophisticated explorations
of these dynamical issues would be possible.  

\acknowledgments
Acknowledgements: We would like to thank M. Pahre for many valuable discussions, M. Franx 
for help identifying the cluster galaxies, and C. Fassnacht and L. Lubin for early access
to their lens redshift measurements.  
Support for the CASTLES project was provided by NASA through grant numbers
GO-7495 and GO-7887 from the Space Telescope Science Institute, which is 
operated by the Association of Universities for Research in Astronomy, Inc.
CSK, EEF, JL, BAM and JAM were also supported by the Smithsonian Institution.
CSK and CRK were also supported by the NASA Astrophysics Theory Program grant 
NAG5-4062.  HWR is also supported by a Fellowship from the Alfred P. Sloan 
Foundation.

\def\mc#1{\multicolumn{1}{c}{#1}}

\vfill\eject
%\input tables1and2.tex
% TABLE 1: SUMMARY OF OBSERVATIONS
%\newpage
\begin{deluxetable}{llrcl}
\label{tab-obslog}
\tablewidth{0pt}
\tablecolumns{6}
\tablecaption{Summary of Lens Observations}
\scriptsize
\tablehead{ \mc{Target}  & Camera/Filter &\mc{Time} &Date &\mc{Source} }
%%%
\startdata
0047-2808        & WFPC2/F555W & 9500  & 99.01.07 & GO-6560, Warren \\
Q0142--100=UM673 & NIC2/F160W  & 2560  & 97.08.15 & Leh\'ar et al. 2000 \\ %GO-7495, CASTLES \\
                 & WFPC2/F675W & 2500  & 94.11.22 & Keeton et al. 1998 \\ %GO-5505, Falco \\
                 & WFPC2/F555W & 1200  & 94.11.22 & Keeton et al. 1998 \\ %GO-5505, Falco \\
MG0414+0534      & NIC2/F205W  &  640  & 97.08.14 & CASTLES \\ %GO-7338, CASTLES \\
                 & NIC2/F160W  &10048  & 98.02.13 & CASTLES \\ %GO-7338, CASTLES \\
                 & NIC2/F110W  & 1792  & 97.08.14 & CASTLES \\ %GO-7338, CASTLES \\
                 & WFPC2/F814W &10500  & 94.11.08 & Falco, Leh\'ar \& Shapiro 1997 \\ %GO-5505, Falco \\
                 & WFPC2/F675W & 8100  & 94.11.08 & Falco, Leh\'ar \& Shapiro 1997 \\ %GO-5505, Falco \\
B0712+472        & NIC1/F160W  & 5248  & 97.08.24 & Jackson et al. 1998a \\ %GO-7255, Jackson \\
                 & WFPC2/F814W & 1000  & 96.01.29 & Jackson et al. 1998b \\ %GO-5908, Jackson \\
                 & WFPC2/F555W &  800  & 96.01.29 & Jackson et al. 1998b \\ %GO-5908, Jackson \\
RXJ0911+0551     & NIC2/F160W  & 2560  & 98.10.18 & CASTLES \\ %GO-7887, CASTLES \\
                 & WFPC2/F814W & 1600  & 00.03.02 & CASTLES \\ %GO-8175, CASTLES \\
FBQ0951+2635     & NIC2/F160W  & 2560  & 98.03.19 & CASTLES \\ %GO-7887, CASTLES \\
                 & WFPC2/F814W & 1600  & 99.11.11 & CASTLES \\ %GO-8175, CASTLES \\
                 & WFPC2/F555W & 1600  & 99.10.08 & CASTLES \\ %GO-7495, CASTLES \\
BRI0952--0115    & NIC2/F160W  & 5120  & 97.10.17 & Leh\'ar et al. 2000 \\ %GO-7495, CASTLES \\
                 & WFPC2/F675W & 5400  & 94.10.22 & Keeton et al. 1998 \\ %GO-5505, Falco \\
Q0957+561        & NIC2/F160W  & 2816  & 98.05.30 & CASTLES \\ %GO-7887, CASTLES \\
                 & WFPC2/F814W & 2620  & 95.11.19 & Bernstein et al. 1997 \\ %GO-5979, Rhee \\
                 & WFPC2/F555W &32200  & 95.11.19 & Bernstein et al. 1997 \\ %GO-5979, Rhee \\
LBQS1009--0252   & NIC2/F160W  & 2560  & 97.11.15 & Leh\'ar et al. 2000 \\ %GO-7495, CASTLES \\
                 & WFPC2/F814W & 2600  & 99.01.01 & Leh\'ar et al. 2000 \\ %GO-6790
                 & WFPC2/F555W & 1600  & 99.01.01 & Leh\'ar et al. 2000 \\ %GO-6790
Q1017--207=J03.13& NIC2/F160W  & 2560  & 97.11.14 & Leh\'ar et al. 2000 \\ %GO-7495, CASTLES \\
                 & WFPC2/F814W & 2300  & 95.11.28 & GO-5958, Surdej \\
FSC10214+472     & NIC2/F205W  &  384  & 97.10.27 & Evans et al. 1999 \\ %GO-7229, Scoville \\
                 & NIC2/F110W  &  384  & 97.10.27 & Evans et al. 1999\\ %GO-7229, Scoville \\
                 & WFPC2/F814W & 6600  & 94.12.10 & Eisenhardt et al. 1996 \\ %GO-5710, Soifer \\
B1030+071        & NIC1/F160W  & 2624  & 97.11.20 & Leh\'ar et al. 2000 \\ %GO-7255, Jackson \\
                 & WFPC2/F814W & 1000  & 97.02.03 & Xanthopoulos et al. 1998 \\ %GO-6629, Jackson \\
                 & WFPC2/F555W & 1000  & 97.02.03 & Xanthopoulos et al. 1998 \\ %GO-6629, Jackson \\
\hline
\tablebreak
HE1104--1805     & NIC2/F160W  & 2560  & 97.11.22 & Leh\'ar et al. 2000 \\ %GO-7495, CASTLES \\
                 & WFPC2/F814W & 1000  & 95.11.19 & Remy et al. 1998 \\ %GO-5958, Surdej \\
                 & WFPC2/F555W & 1600  & 00.02.04 & CASTLES \\ %GO-7495, CASTLES \\
PG1115+080       & NIC2/F160W  & 2560  & 97.11.17 & Impey et al. 1998 \\ %GO-7495, CASTLES \\
                 & WFPC2/F814W & 4400  & 97.05.17 & GO-6555, Schechter \\
                 & WFPC2/F555W & 3200  & 99.03.31 & CASTLES \\ %GO-7495, CASTLES \\
B1127+385        & NIC1/F160W  & 2624  & 98.04.10 & GO-7873, Wilkinson \\
                 & WFPC2/F814W & 1000  & 96.06.21 & Koopmans et al. 1999 \\ %GO-6629, Jackson \\
MG1131+0456      & NIC2/F160W  & 5120  & 98.01.05 & Kochanek et al. 1999 \\ %GO-7495, CASTLES \\
                 & WFPC2/F814W &10500  & 95.04.18 & Kochanek et al. 1999 \\ %GO-5505, CASTLES \\
                 & WFPC2/F675W & 8100  & 95.04.18 & Kochanek et al. 1999 \\ %GO-5505, CASTLES \\
HST12531--921    & NIC2/F160W  & 5120  & 98.02.14 & CASTLES \\ %GO-7495, CASTLES \\
                 & WFPC2/F814W & 8400  & 95.02.15 & Ratnatunga et al. 1995 \\ %GO-5369, MDS \\
                 & WFPC2/F606W & 5400  & 95.02.15 & Ratnatunga et al. 1995 \\ %GO-5369, MDS \\
HST14113+521     & WFPC2/F702W &12600  & 94.07.25 & Fischer, Schade \& Barientos, 1998\\ %GO-5738, Conti \\
HST14176+522     & NIC2/F160W  & 5632  & 98.05.28 & CASTLES \\ %GO-7495, CASTLES \\
                 & WFPC2/F814W & 4400  & 94.03.11 & Ratnatunga et al. 1995\\ %GO-5090, MDS \\
                 & WFPC2/F606W & 2800  & 94.03.11 & Ratnatunga et al. 1995\\ %GO-5090, MDS \\
B1422+231        & NIC2/F160W  & 5120  & 98.02.27 & CASTLES \\ %GO-7495, CASTLES \\
                 & WFPC2/F791W & 4200  & 99.02.06 & GO-6652, Impey \\
                 & WFPC2/F555W & 4200  & 99.02.06 & GO-6652, Impey \\
SBS1520+530      & NIC2/F160W  & 2816  & 98.07.20 & CASTLES \\ %GO-7887, CASTLES \\
                 & WFPC2/F814W & 1600  & 99.11.09 & CASTLES \\ %GO-8175, CASTLES \\
MG1549+3047      & NIC2/F205W  &  704  & 97.08.17 & CASTLES \\ %GO-7495, CASTLES \\
                 & NIC2/F160W  & 1536  & 97.08.17 & CASTLES \\ %GO-7495, CASTLES \\
                 & WFPC2/F814W &  560  & 99.05.20 & CASTLES \\ %GO-7495, CASTLES \\
                 & WFPC2/F555W &  560  & 99.05.20 & CASTLES \\ %GO-7495, CASTLES \\
B1608+656        & NIC2/F160W  & 2816  & 97.09.29 & CASTLES \\ %GO-7495, CASTLES \\
                 & WFPC2/F814W & 2400  & 96.04.07 & Jackson et al. 1998a \\ %GO-5908, Jackson \\
                 & WFPC2/F555W & 1500  & 96.04.07 & Jackson et al. 1998a \\ %GO-5908, Jackson \\
\hline
\tablebreak
MG1654+1346      & NIC2/F160W  & 2560  & 97.10.12 & CASTLES \\ %GO-7495, CASTLES \\
                 & WFPC2/F814W &10500  & 96.01.19 & Keeton, Kochanek \& Falco 1998 \\ %GO-5505, Falco \\
                 & WFPC2/F675W & 5826  & 96.01.18 & Keeton, Kochanek \& Falco 1998 \\ %GO-5505, Falco \\
B1938+666        & NIC2/F160W  & 5632  & 97.10.07 & CASTLES \\ %GO-7495, CASTLES \\
                 & WFPC2/F814W & 3000  & 99.04.24 & CASTLES \\ %GO-7495, CASTLES \\
                 & WFPC2/F555W & 2800  & 99.04.24 & CASTLES \\ %GO-7495, CASTLES \\
MG2016+112       & NIC2/F160W  & 5120  & 97.10.30 & CASTLES \\ %GO-7495, CASTLES \\
                 & WFPC2/F814W & 2600  & 99.05.14 & GO-6543, Lawrence \\
                 & WFPC2/F555W & 2600  & 99.05.14 & GO-6543, Lawrence \\
B2045+265        & NIC1/F160W  & 2624  & 97.07.14 & Fassnacht et al. 1999 \\ %GO-7255, Jackson \\
HE2149--2745     & NIC2/F160W  & 2560  & 98.09.04 & CASTLES \\ %GO-7887, CASTLES \\
                 & WFPC2/F814W & 1600  & 99.10.24 & CASTLES \\ %GO-8175, CASTLES \\
Q2237+030        & NIC2/F205W  &  704  & 97.10.11 & CASTLES \\ %GO-7495, CASTLES \\
                 & NIC2/F160W  & 1532  & 97.10.11 & CASTLES \\ %GO-7495, CASTLES \\
                 & WFPC2/F814W &  120  & 99.10.20 & CASTLES \\ %GO-8252, CASTLES \\
                 & WFPC2/F555W & 1600  & 95.06.23 & GO-5236, Westphal \\
B2319+052        & NIC1/F160W  & 2624  & 98.05.30 & GO-7873, Wilkinson \\
\enddata
\tablecomments{
   The exposure time is in seconds.  The Source entry is either the
   first published discussion of the data or the HST program ID and PI
   for unpublished data.    
   }
\end{deluxetable}

\vfill\eject

\begin{deluxetable}{lccccrcl}
\label{tab-obslog}
\tablewidth{0pt}
\tablecolumns{9}
\tablecaption{Summary of Cluster Observations}
\scriptsize
\tablehead{ \mc{Cluster} &\mc{$z_c$} &\mc{$E_{Gal}$} &\mc{\#} & Filter &\mc{Time} 
     &Date &Source }
%%%
\startdata
A665           &0.18 &0.045 &1     & F814W & 4800 & 94.12.02 & GO-5458, Franx        \\
               &     &      &1     & F606W & 5100 & 94.12.02 & GO-5458, Franx         \\
               &     &      &2     & F814W & 4400 & 94.10.31 & GO-5458, Franx         \\
               &     &      &2     & F606W & 4400 & 94.10.31 & GO-5458, Franx         \\
\\
A2390          &0.23 &0.113 &1     & F814W &10500 & 94.12.10 & GO-5352, Fort          \\
               &     &      &1     & F555W & 8400 & 94.12.10 & GO-5352, Fort          \\
\\
CL1358+62      &0.33 &0.024 &1     & F814W & 3600 & 96.02.12 &Kelson et al. 1997      \\
               &     &      &1     & F606W & 3600 & 96.02.12 &Kelson et al. 1997      \\
\\
A370           &0.37 &0.032 &1     & F675W & 5600 & 95.12.02 &Ziegler et al. 1999     \\
               &     &      &2     & F814W &12600 & 95.01.12 &Smail et al. 1997       \\
               &     &      &2     & F555W & 8000 & 95.01.12 &Smail et al. 1997       \\
\\
A851           &0.41 &0.015 &1     & F702W & 4400 & 97.04.23 &GO-6480, Dressler      \\
               &     &      &2     & F702W &21000 & 94.01.10 &Dressler et al. 1994   \\
               &     &      &3     & F814W &12600 & 94.04.18 &Smail et al. 1997      \\
               &     &      &3     & F555W & 8000 & 94.04.18 &Smail et al. 1997      \\
\\
MS0015+16      &0.55 &0.056 &1     & F814W &16800 & 94.12.11 &Smail et al. 1997       \\
               &     &      &1     & F555W &12600 & 94.12.11 &Smail et al. 1997       \\
\\
MS2053--04     &0.58 &0.084 &1     & F814W & 2100 & 97.12.13 &Kelson et al. 1997     \\
               &     &      &1     & F702W & 2400 & 95.10.23 &Kelson et al. 1997     \\
               &     &      &2     & F814W & 1100 & 97.12.11 &Kelson et al. 1997     \\
               &     &      &3     & F814W & 2100 & 97.12.13 &Kelson et al. 1997     \\
\\
MS1054--03     &0.83 &0.024 &1     & F814W &15600 & 96.03.13 &Donahue et al. 1998  \\
               &     &      &1     & F606W & 6500 & 98.05.26 &van Dokkum et al. 1999  \\
\\
\enddata
\tablecomments{
   The ``\#" entry is an arbitrary number to indicate which images overlap. The exposure time is in
   seconds.  $E_{Gal}=E(B-V)$ is the foreground Galactic extinction from Schlegel, Finkbeiner \& Davis (1998).  
   The source of the photometry is either the first published paper discussing the observations
   or the program identification. 
   %for unpublished data.    
   }
\end{deluxetable}

\vfill\eject
\begin{deluxetable}{rccrcccccc}
%\scriptsize
\tablecaption{ Lens Galaxy Photometric Data }
\tablewidth{0pt}
\tablehead{\mc{Lens} &$E_{Gal}$ &$\Delta\theta$ &\mc{$\log(r_e/'')$} &$\mu_e$        &$\sigma_{FP}$ &Filter 1 &color &Filter 2  \\
                     &(mag)    &($''$)         &                    &(mag/asec$^2$) &                   &         &(mag) &         }
\startdata
0047-2808    & 0.016 &$ 2.70$         &$ -0.04\pm  0.02$ &$ 22.45\pm  0.08$ &$  0.01$&F555W \\
Q0142-100    & 0.031 &$ 2.24$         &$ -0.29\pm  0.02$ &$ 17.17\pm  0.09$ &$  0.03$&F160W   
        &$  4.18\pm  0.02 $&F555W \\
&&&&&&  &$  2.72\pm  0.01 $&F675W \\
MG0414+0534  & 0.303 &$ 2.38$         &$ -0.11\pm  0.08$ &$ 18.98\pm  0.23$ &$  0.02$&F160W   
        &$  1.67\pm  0.03 $&F110W \\
&&&&&&  &$  0.84\pm  0.12 $&F205W \\
&&&&&&  &$  5.04\pm  0.13 $&F675W \\
&&&&&&  &$  3.37\pm  0.05 $&F814W \\
B0712+472    & 0.113 &$ 1.42$         &$ -0.44\pm  0.06$ &$ 16.95\pm  0.14$ &$  0.05$&F160W   
        &$  4.47\pm  0.06 $&F555W \\
&&&&&&  &$  2.34\pm  0.07 $&F814W \\
RXJ0911+0551 & 0.045 &$ 2.21$         &$ -0.17\pm  0.04$ &$ 19.10\pm  0.14$ &$  0.02$&F160W   
        &$  2.54\pm  0.09 $&F814W \\
FBQ0951+2635 & 0.022 &$ 1.11$         &$ -0.78\pm  0.10$ &$ 15.96\pm  0.32$ &$  0.07$&F160W   
        &$  3.16\pm  0.04 $&F555W \\
&&&&&&  &$  1.81\pm  0.03 $&F814W \\
BRI0952-0115 & 0.063 &$ 1.00$         &$ -1.00\pm  0.12$ &$ 15.97\pm  0.45$ &$  0.09$&F160W   
        &$  3.13\pm  0.03 $&F675W \\
Q0957+561    & 0.009 &$ 6.26$         &$  0.30\pm  0.04$ &$ 18.63\pm  0.11$ &$  0.03$&F160W   
        &$  3.91\pm  0.06 $&F555W \\
&&&&&&  &$  1.98\pm  0.03 $&F814W \\
LBQS1009-025 & 0.034 &$ 1.54$         &$ -0.71\pm  0.08$ &$ 17.69\pm  0.28$ &$  0.05$&F160W   
        &$  5.46\pm  0.22 $&F555W \\
&&&&&&  &$  2.63\pm  0.04 $&F814W \\
Q1017-207    & 0.046 &$ 0.85$         &$ -0.52\pm  0.02$ &$ 18.64\pm  0.06$ &$  0.06$&F160W   
        &$  2.56\pm  0.48 $&F814W \\
FSC10214+472 & 0.012 &$ 1.59$         &$  0.05\pm  0.05$ &$ 22.67\pm  0.12$ &$  0.03$&F814W   
        &$  1.11\pm  0.36 $&F110W \\
&&&&&&  &$  3.36\pm  0.41 $&F205W \\
B1030+071    & 0.022 &$ 1.56$         &$ -0.37\pm  0.03$ &$ 20.32\pm  0.09$ &$  0.01$&F814W   
        &$  2.51\pm  0.21 $&F160W \\
&&&&&&  &$  2.56\pm  0.04 $&F555W \\
HE1104-1805  & 0.056 &$ 3.19$         &$ -0.19\pm  0.13$ &$ 18.49\pm  0.38$ &$  0.03$&F160W   
        &$  5.67\pm  0.50 $&F555W \\
&&&&&&  &$  2.54\pm  0.10 $&F814W \\
PG1115+080   & 0.041 &$ 2.29$         &$ -0.33\pm  0.02$ &$ 17.00\pm  0.08$ &$  0.02$&F160W   
        &$  4.08\pm  0.04 $&F555W \\
&&&&&&  &$  2.26\pm  0.02 $&F814W \\
B1127+385    & 0.027 &$ 0.70$         &$ -1.01\pm  0.21$ &$ 16.97\pm  0.56$ &$  0.07$&F160W   
        &$  2.68\pm  0.56 $&F814W \\
MG1131+0456  & 0.036 &$ 2.10$         &$ -0.24\pm  0.05$ &$ 19.42\pm  0.17$ &$  0.04$&F160W   
        &$  3.85\pm  0.06 $&F675W \\
&&&&&&  &$  2.59\pm  0.04 $&F814W \\
\hline     
\tablebreak
HST12531-291 & 0.079 &$ 1.09$         &$ -0.85\pm  0.03$ &$ 17.22\pm  0.11$ &$  0.04$&F160W   
        &$  4.33\pm  0.03 $&F606W \\
&&&&&&  &$  2.33\pm  0.03 $&F814W \\
HST14113+521 & 0.016 &$ 1.72$         &$ -0.32\pm  0.05$ &$ 20.87\pm  0.17$ &$  0.04$&F702W \\
HST14176+522 & 0.007 &$ 2.84$         &$ -0.15\pm  0.05$ &$ 18.76\pm  0.14$ &$  0.03$&F160W   
        &$  4.36\pm  0.08 $&F606W \\
&&&&&&  &$  2.20\pm  0.09 $&F814W \\
B1422+231    & 0.048 &$ 1.56$         &$ -0.50\pm  0.13$ &$ 17.07\pm  0.45$ &$  0.06$&F160W   
        &$  4.23\pm  0.05 $&F555W \\
&&&&&&  &$  2.09\pm  0.06 $&F791W \\
SBS1520+530  & 0.016 &$ 1.59$         &$ -0.46\pm  0.04$ &$ 17.52\pm  0.12$ &$  0.02$&F160W   
        &$  2.32\pm  0.13 $&F814W \\
MG1549+3047  & 0.029 &$ 1.70$         &$ -0.06\pm  0.02$ &$ 16.39\pm  0.04$ &$  0.02$&F160W   
        &$  0.68\pm  0.01 $&F205W \\
&&&&&&  &$  3.51\pm  0.02 $&F555W \\
&&&&&&  &$  2.02\pm  0.02 $&F814W \\
B1608+656    & 0.031 &$ 2.27$         &$ -0.19\pm  0.07$ &$ 17.78\pm  0.23$ &$  0.02$&F160W   
        &$  4.48\pm  0.23 $&F555W \\
&&&&&&  &$  2.18\pm  0.35 $&F814W \\
MG1654+1346  & 0.061 &$ 2.10$         &$ -0.05\pm  0.02$ &$ 17.57\pm  0.04$ &$  0.02$&F160W   
        &$  2.72\pm  0.01 $&F675W \\
&&&&&&  &$  2.07\pm  0.02 $&F814W \\
B1938+666    & 0.121 &$ 1.00$         &$ -0.16\pm  0.04$ &$ 19.86\pm  0.12$ &$  0.04$&F160W   
        &$  5.78\pm  0.84 $&F555W \\
&&&&&&  &$  2.79\pm  0.08 $&F814W \\
MG2016+112   & 0.235 &$ 3.26$         &$ -0.68\pm  0.03$ &$ 17.05\pm  0.12$ &$  0.04$&F160W   
        &$  7.39\pm  0.08 $&F555W \\
&&&&&&  &$  3.47\pm  0.01 $&F814W \\
B2045+265    & 0.235 &$ 2.28$         &$ -0.42\pm  0.13$ &$ 18.14\pm  0.44$ &$  0.09$&F160W \\
HE2149-2745  & 0.072 &$ 1.70$         &$ -0.30\pm  0.02$ &$ 18.10\pm  0.05$ &$  0.05$&F160W   
        &$  1.95\pm  0.03 $&F814W \\
Q2237+030    & 0.071 &$ 1.76$         &$  0.55\pm  0.11$ &$ 17.06\pm  0.29$ &$  0.07$&F160W   
        &$  0.44\pm  0.26 $&F205W \\
&&&&&&  &$  3.29\pm  0.28 $&F555W \\
&&&&&&  &$  2.44\pm  0.01 $&F675W \\
&&&&&&  &$  2.26\pm  0.01 $&F814W \\
B2319+052    & 0.064 &$ 1.36$         &$ -0.65\pm  0.02$ &$ 16.92\pm  0.05$ &$  0.03$&F160W \\
\enddata
\tablecomments{ For each lens, $E_{Gal}=E(B-V)$ is the foreground Galactic extinction from Schlegel, Finkbeiner \& 
  Davis (1998) and $\Delta\theta$ is the image separation.    For the reference filter (Filter 1) we present values 
  and uncertainties for the logarithm of the intermediate axis effective radius, $\log(r_e/'')$, the mean 
  surface brightness inside the effective radius, $\mu_e$, and the uncertainty $\sigma_{FP}$ for the 
  variable combination $\mu_e-3.03\log(r_e)$ appearing in the fundamental plane equations (see eqn. (2)).  
  We then provide the colors measured between Filter 1 and Filter 2 (blue minus red), with one
  color per line of the table.  The magnitudes and colors are {\it not} corrected for the foreground
  extinction, but an $R_V=3.1$ extinction curve is used to correct the magnitudes in all calculations 
  and figures. }
\end{deluxetable}

\vfill\eject
\vspace{-0.50in}
\begin{deluxetable}{rrcrccccc}
%\scriptsize
\tablecaption{ Cluster Galaxy Photometric Data }
\tablewidth{0pt}
\tablehead{
  \mc{Cluster} &\mc{Galaxy} &\mc{$\log(\sigma_c/\hbox{km s}^{-1})$}   &\mc{$\log(r_e/'')$} &$\mu_e$        &Filter 1 &color &Filter 2 \\
               &            &                                         &                    &(mag/asec$^2$) &         &(mag) &         }
\startdata
A665&   3            &$2.439\pm0.013$ &$  0.34$ &$ 19.78$ &F814W   
        &$  1.05\pm  0.05 $&F606W \\
    &  15            &$2.414\pm0.021$ &$  0.18$ &$ 19.70$ &F814W   
        &$  1.05\pm  0.05 $&F606W \\
    &  26            &$2.355\pm0.024$ &$  0.03$ &$ 19.28$ &F814W   
        &$  1.02\pm  0.05 $&F606W \\
    &  42            &$2.395\pm0.018$ &$  0.02$ &$ 19.21$ &F814W   
        &$  1.11\pm  0.05 $&F606W \\
    &  57            &$2.325\pm0.021$ &$ -0.18$ &$ 18.73$ &F814W   
        &$  1.01\pm  0.05 $&F606W \\
    &  61            &$2.358\pm0.012$ &$ -0.15$ &$ 18.93$ &F814W   
        &$  1.13\pm  0.05 $&F606W \\
    &  77            &$2.173\pm0.030$ &$  0.11$ &$ 20.39$ &F814W   
        &$  1.06\pm  0.05 $&F606W \\
    &  80            &$2.276\pm0.028$ &$ -0.18$ &$ 19.12$ &F814W   
        &$  0.99\pm  0.05 $&F606W \\
A2390&   6           &$2.313\pm0.017$ &$  0.01$ &$ 19.67$ &F814W   
        &$  1.65\pm  0.05 $&F555W \\
     &   7           &$2.277\pm0.019$ &$  0.17$ &$ 20.16$ &F814W   
        &$  1.68\pm  0.05 $&F555W \\
     &   9           &$2.371\pm0.011$ &$ -0.21$ &$ 18.63$ &F814W   
        &$  1.64\pm  0.05 $&F555W \\
     &  10           &$2.250\pm0.020$ &$ -0.22$ &$ 19.33$ &F814W   
        &$  1.76\pm  0.05 $&F555W \\
     & 138           &$2.029\pm0.017$ &$ -0.13$ &$ 20.87$ &F814W   
        &$  1.66\pm  0.05 $&F555W \\
CL1358& 236          &$2.220\pm0.029$ &$ -0.24$ &$ 19.69$ &F814W   
        &$  1.26\pm  0.05 $&F606W \\
      & 256          &$2.436\pm0.011$ &$ -0.01$ &$ 19.43$ &F814W   
        &$  1.17\pm  0.05 $&F606W \\
      & 269          &$2.534\pm0.013$ &$ -0.08$ &$ 19.27$ &F814W   
        &$  1.28\pm  0.05 $&F606W \\
      & 298          &$2.447\pm0.012$ &$ -0.13$ &$ 19.31$ &F814W   
        &$  1.21\pm  0.05 $&F606W \\
      & 375          &$2.479\pm0.016$ &$  0.39$ &$ 20.96$ &F814W   
        &$  1.27\pm  0.05 $&F606W \\
      & 408          &$2.423\pm0.028$ &$ -0.40$ &$ 19.02$ &F814W   
        &$  1.17\pm  0.05 $&F606W \\
      & 454          &$2.233\pm0.015$ &$  0.19$ &$ 21.08$ &F814W   
        &$  1.22\pm  0.05 $&F606W \\
      & 470          &$2.267\pm0.014$ &$ -0.04$ &$ 19.91$ &F814W   
        &$  1.15\pm  0.05 $&F606W \\
A370&   1            &$2.519\pm0.005$ &$  0.33$ &$ 21.51$ &F675W \\
    &   2            &$2.404\pm0.007$ &$  0.95$ &$ 23.42$ &F675W \\
    &  10            &$2.291\pm0.009$ &$  0.15$ &$ 21.66$ &F675W \\
    &  24            &$2.399\pm0.007$ &$ -0.10$ &$ 20.51$ &F675W \\
    &  28            &$2.345\pm0.008$ &$ -0.17$ &$ 19.92$ &F814W   
        &$  1.90\pm  0.05 $&F555W \\
    &  41            &$2.467\pm0.009$ &$ -0.29$ &$ 19.39$ &F814W   
        &$  2.07\pm  0.05 $&F555W \\
    &  67            &$2.207\pm0.011$ &$  0.01$ &$ 21.05$ &F814W   
        &$  1.96\pm  0.05 $&F555W \\
    &  77            &$1.976\pm0.019$ &$ -0.04$ &$ 21.71$ &F675W \\
    &  79            &$2.227\pm0.011$ &$ -0.34$ &$ 20.28$ &F675W \\
A851&  23            &$2.273\pm0.019$ &$ -0.19$ &$ 20.31$ &F702W \\
    &  57            &$2.299\pm0.018$ &$ -0.39$ &$ 19.91$ &F702W \\
    &  69            &$2.288\pm0.014$ &$ -0.24$ &$ 20.43$ &F814W   
        &$  1.86\pm  0.05 $&F555W \\
    & 102            &$2.173\pm0.018$ &$ -0.63$ &$ 19.78$ &F702W \\
    & 111            &$1.766\pm0.031$ &$ -0.26$ &$ 21.36$ &F702W \\
\hline
\tablebreak
MS0015&   2          &$2.412\pm0.018$ &$  1.01$ &$ 23.84$ &F814W   
        &$  2.48\pm  0.05 $&F555W \\
      &   7          &$2.291\pm0.009$ &$ -0.29$ &$ 19.97$ &F814W   
        &$  2.59\pm  0.05 $&F555W \\
      &  13          &$2.422\pm0.017$ &$ -0.39$ &$ 19.69$ &F814W   
        &$  2.40\pm  0.05 $&F555W \\
      &  56          &$2.337\pm0.017$ &$ -1.50$ &$ 15.68$ &F814W   
        &$  2.48\pm  0.05 $&F555W \\
MS2053& 197          &$2.504\pm0.025$ &$  0.20$ &$ 21.59$ &F814W   
        &$  0.83\pm  0.05 $&F702W \\
      & 311          &$2.348\pm0.049$ &$ -0.42$ &$ 20.45$ &F814W   
        &$  0.78\pm  0.05 $&F702W \\
      & 422          &$2.120\pm0.060$ &$ -0.51$ &$ 20.05$ &F814W   
        &$  0.74\pm  0.05 $&F702W \\
      & 432          &$2.207\pm0.054$ &$ -0.30$ &$ 20.96$ &F814W   
        &$  0.73\pm  0.05 $&F702W \\
      & 551          &$2.336\pm0.038$ &$ -0.66$ &$ 19.56$ &F814W   
        &$  0.79\pm  0.05 $&F702W \\
MS1054&1294          &$2.500\pm0.029$ &$ -0.18$ &$ 21.35$ &F814W   
        &$  2.18\pm  0.05 $&F606W \\
      &1359          &$2.352\pm0.037$ &$ -0.52$ &$ 20.39$ &F814W   
        &$  2.43\pm  0.05 $&F606W \\
      &1405          &$2.413\pm0.035$ &$ -0.02$ &$ 21.87$ &F814W   
        &$  2.25\pm  0.05 $&F606W \\
      &1457          &$2.322\pm0.050$ &$ -0.24$ &$ 21.42$ &F814W   
        &$  2.10\pm  0.05 $&F606W \\
      &1484          &$2.519\pm0.026$ &$  0.19$ &$ 22.24$ &F814W   
        &$  2.43\pm  0.05 $&F606W \\
      &1567          &$2.417\pm0.045$ &$ -0.33$ &$ 21.09$ &F814W   
        &$  2.17\pm  0.05 $&F606W \\
\enddata
\tablecomments{The variables are the same as in Table 3 except for the velocity dispersion $\sigma_c$ 
   replacing the image separation $\Delta\theta$. The velocity dispersions for A665, A2390, A370, A851 and MS0015+16
   are from Pahre et al. (1999a), those for CL1358+62 and MS2053--04 are from Kelson et al. (1997),
   and those for MS1054--03 are from van Dokkum et al. (1998).}
\end{deluxetable}

\vfill\eject
\vspace{-0.50in}
\begin{deluxetable}{ccccccc}
%\footnotesize
\tablecaption{ Lens Galaxy Redshift Estimates }
\tablewidth{0pt}
\tablehead{Lens &$z_l$ &$z_s$ &$z_{FP}$ &$z_{min}$ &$z_{max}$ &$N_{filt}$}
\startdata
0047-2808    &  0.48 &  3.60 & 0.39  & 0.35  & 0.48  & 1\\
Q0142-100    &  0.49 &  2.72 & 0.36  & 0.33  & 0.38  & 3\\
MG0414+0534  &  0.96 &  2.64 & 0.89  & 0.76  & 0.94  & 5\\
B0712+472    &  0.41 &  1.34 & 0.33  & 0.31  & 0.35  & 3\\
RXJ0911+0551 &  0.77 &  2.80 & 0.82  & 0.74  & 0.92  & 2\\
FBQ0951+2635 &       &  1.24 & 0.21  & 0.18  & 0.23  & 3\\
BRI0952-0115 &       &  4.50 & 0.41  & 0.36  & 0.46  & 2\\
Q0957+561    &  0.36 &  1.41 & 0.37  & 0.34  & 0.39  & 3\\
LBQS1009-025 &       &  2.74 & 0.88  & 0.77  & 0.92  & 3\\
Q1017-207    &       &  2.55 & 0.78  & 0.73  & 0.87  & 2\\
FSC10214+472 &       &  2.29 & 0.78  & 0.68  & 0.81  & 3\\
B1030+071    &  0.60 &  1.54 & 0.54  & 0.51  & 0.68  & 3\\
HE1104-1805  &       &  2.32 & 0.73  & 0.69  & 0.76  & 3\\
PG1115+080   &  0.31 &  1.72 & 0.29  & 0.27  & 0.35  & 3\\
B1127+385    &       &       & 0.78  & 0.62  & 1.03  & 2\\
MG1131+0456  &  0.84 &       & 0.95  & 0.86  & 1.01  & 3\\
HST12531-291 &       &       & 0.63  & 0.60  & 0.83  & 3\\
HST14113+521 &  0.46 &  2.81 & 0.51  & 0.44  & 0.57  & 1\\
HST14176+522 &  0.81 &  3.40 & 0.71  & 0.66  & 0.79  & 3\\
B1422+231    &  0.34 &  3.62 & 0.37  & 0.35  & 0.39  & 3\\
SBS1520+530  &       &  1.86 & 0.52  & 0.44  & 0.65  & 2\\
MG1549+3047  &  0.11 &       & 0.12  & 0.10  & 0.14  & 4\\
B1608+656    &  0.63 &  1.39 & 0.45  & 0.43  & 0.48  & 3\\
MG1654+1346  &  0.25 &  1.74 & 0.25  & 0.22  & 0.29  & 3\\
B1938+666    &  0.88 &       & 0.84  & 0.81  & 1.04  & 3\\
MG2016+112   &  1.01 &  3.27 & 0.99  & 0.96  & 1.01  & 3\\
B2045+265    &  0.87 &  1.28 & 0.68  & 0.56  & 0.76  & 1\\
HE2149-2745  &       &  2.03 & 0.43  & 0.37  & 0.50  & 2\\
Q2237+030    &  0.04 &  1.69 & 0.02  & 0.02  & 0.03  & 5\\
B2319+052    &  0.62 &       & 0.61  & 0.27  & 0.88  & 1\\
\enddata
\tablecomments{ $N_{filt}$ is the number of filters available for the measurement.
  All examples with $N_{filt}=1$ except HST14113+521 are H band observations.  The 
  redshift uncertainties are the formal uncertainties defined by the region where
  $\Delta \chi^2 < 4$.  The actual accuracy is better characterized by the scatter
  observed in Figure 5.  
  }
\end{deluxetable}

\vfill\eject
\vspace{-0.50in}
\begin{deluxetable}{ccccrrr}
%\scriptsize
\tablecaption{ E+K Corrections }
\tablewidth{0pt}
\tablehead{ Data &Band &N &$\langle z \rangle$ &\mc{$\Omega_0=1$ Flat    } &\mc{$\Omega_0=0.3$ Open  } &\mc{$\Omega_0=0.3$ Flat  }\\
                 &     &  &                    &\mc{$\langle e+k \rangle$} &\mc{$\langle e+k \rangle$} &\mc{$\langle e+k \rangle$} }
\startdata
Lens 1 % 1
&V &4 &$ 0.11 \pm  0.09$ &$ 0.21 \pm  0.12$ 
&$ 0.19 \pm  0.10$ 
&$ 0.13 \pm  0.08$  \\ 
&  &8 &$ 0.39 \pm  0.09$ &$ 0.71 \pm  0.12$ 
&$ 0.62 \pm  0.12$ 
&$ 0.41 \pm  0.13$  \\ 
&  &2 &$ 0.62 \pm  0.01$ &$ 1.21 \pm  0.43$ 
&$ 1.06 \pm  0.44$ 
&$ 0.80 \pm  0.44$  \\ 
&  &5 &$ 0.90 \pm  0.07$ &$ 2.34 \pm  0.22$ 
&$ 2.16 \pm  0.21$ 
&$ 1.91 \pm  0.20$  \\ 
&I &3 &$ 0.13 \pm  0.09$ &$ 0.08 \pm  0.13$ 
&$ 0.07 \pm  0.13$ 
&$ 0.01 \pm  0.15$  \\ 
&  &5 &$ 0.36 \pm  0.07$ &$ 0.19 \pm  0.12$ 
&$ 0.12 \pm  0.11$ 
&$-0.06 \pm  0.12$  \\ 
&  &2 &$ 0.62 \pm  0.01$ &$-0.01 \pm  0.30$ 
&$-0.14 \pm  0.30$ 
&$-0.39 \pm  0.30$  \\ 
&  &6 &$ 0.88 \pm  0.08$ &$ 0.49 \pm  0.11$ 
&$ 0.35 \pm  0.12$ 
&$ 0.09 \pm  0.12$  \\ 
&H &5 &$ 0.11 \pm  0.08$ &$-0.20 \pm  0.09$ 
&$-0.15 \pm  0.08$ 
&$-0.18 \pm  0.08$ \\ 
&  &6 &$ 0.36 \pm  0.08$ &$-0.20 \pm  0.17$ 
&$-0.23 \pm  0.17$ 
&$-0.39 \pm  0.18$ \\ 
&  &3 &$ 0.62 \pm  0.01$ &$-0.50 \pm  0.07$ 
&$-0.65 \pm  0.10$ 
&$-0.69 \pm  0.12$ \\ 
&  &9 &$ 0.90 \pm  0.08$ &$-0.32 \pm  0.12$ 
&$-0.45 \pm  0.13$ 
&$-0.69 \pm  0.13$ \\ 
\tableline % 1
Lens 2 % 1
&V &5 &$ 0.14 \pm  0.09$ &$ 0.21 \pm  0.09$ 
&$ 0.19 \pm  0.07$ 
&$ 0.14 \pm  0.06$  \\ 
&  &9 &$ 0.40 \pm  0.09$ &$ 0.74 \pm  0.11$ 
&$ 0.67 \pm  0.12$ 
&$ 0.46 \pm  0.13$  \\ 
&  &2 &$ 0.62 \pm  0.01$ &$ 1.21 \pm  0.43$ 
&$ 1.14 \pm  0.26$ 
&$ 1.23 \pm  0.33$  \\ 
&  &8 &$ 0.89 \pm  0.09$ &$ 2.13 \pm  0.19$ 
&$ 2.12 \pm  0.15$ 
&$ 1.86 \pm  0.17$  \\ 
&I &4 &$ 0.16 \pm  0.09$ &$ 0.07 \pm  0.10$ 
&$ 0.07 \pm  0.09$ 
&$ 0.03 \pm  0.11$  \\ 
&  &5 &$ 0.36 \pm  0.07$ &$ 0.19 \pm  0.12$ 
&$ 0.08 \pm  0.10$ 
&$-0.07 \pm  0.09$  \\ 
&  &4 &$ 0.63 \pm  0.07$ &$-0.13 \pm  0.14$ 
&$-0.16 \pm  0.13$ 
&$-0.18 \pm  0.14$  \\ 
&  &13 &$ 0.88 \pm  0.09$ &$ 0.30 \pm  0.10$ 
&$ 0.31 \pm  0.08$ 
&$ 0.11 \pm  0.08$  \\ 
&H &6 &$ 0.13 \pm  0.09$ &$-0.14 \pm  0.09$ 
&$-0.10 \pm  0.08$ 
&$-0.11 \pm  0.09$ \\ 
&  &7 &$ 0.38 \pm  0.08$ &$-0.26 \pm  0.15$ 
&$-0.25 \pm  0.13$ 
&$-0.36 \pm  0.14$ \\ 
&  &5 &$ 0.63 \pm  0.07$ &$-0.49 \pm  0.09$ 
&$-0.61 \pm  0.07$ 
&$-0.58 \pm  0.07$ \\ 
&  &17 &$ 0.89 \pm  0.08$ &$-0.36 \pm  0.08$ 
&$-0.40 \pm  0.09$ 
&$-0.60 \pm  0.10$ \\ 
\tableline
Cluster % 1
&V &13 &$ 0.20 \pm  0.02$ &$ 0.37 \pm  0.06$ 
&$ 0.33 \pm  0.07$ 
&$ 0.25 \pm  0.07$  \\ 
&  &18 &$ 0.35 \pm  0.02$ &$ 0.67 \pm  0.08$ 
&$ 0.55 \pm  0.08$ 
&$ 0.41 \pm  0.07$  \\ 
&  &4 &$ 0.55 \pm  0.00$ &$ 0.79 \pm  0.20$ 
&$ 0.66 \pm  0.20$ 
&$ 0.52 \pm  0.20$  \\ 
&  &6 &$ 0.83 \pm  0.00$ &$ 2.01 \pm  0.09$ 
&$ 1.83 \pm  0.09$ 
&$ 1.65 \pm  0.10$  \\ 
&I &13 &$ 0.20 \pm  0.02$ &$ 0.04 \pm  0.06$ 
&$ 0.01 \pm  0.06$ 
&$-0.06 \pm  0.06$  \\ 
&  &16 &$ 0.36 \pm  0.04$ &$-0.13 \pm  0.10$ 
&$-0.19 \pm  0.10$ 
&$-0.30 \pm  0.10$  \\ 
&  &14 &$ 0.57 \pm  0.01$ &$-0.12 \pm  0.11$ 
&$-0.24 \pm  0.11$ 
&$-0.37 \pm  0.11$  \\ 
&  &6 &$ 0.83 \pm  0.00$ &$ 0.17 \pm  0.08$ 
&$ 0.01 \pm  0.07$ 
&$-0.14 \pm  0.08$  \\ 
\enddata
\tablecomments{The ``Lens 1'' data include only lenses with known redshifts, while the ``Lens 2'' data include
  all systems using the redshift estimates of \S4.  The E+K corrections are computed in redshift bins with 
  edges at $z=0.25$, $0.50$ and $0.75$.  For each bin, $\langle z \rangle$ gives the mean and 
  dispersion of the redshifts for the $N$ estimates in the bin. The uncertainty in the E+K correction is simply 
  the dispersion of the values in the bin divided by $(N-1)^{1/2}$.  
  }
\end{deluxetable}
\vfill

\end{document}